\documentclass[10pt,a4paper,amsmath,amssymb,aps,pre,notitlepage]{revtex4-1}
\usepackage[caption=false]{subfig}
\usepackage{graphicx,subfig}
\usepackage{dcolumn}
\usepackage{bm}
\usepackage{graphicx,amssymb,color,enumerate,setspace,upgreek,amsmath,appendix,fancyhdr,titlesec,wasysym,mnsymbol,multirow}
\usepackage[usenames,dvipsnames,svgnames,table]{xcolor}

\begin{document}

\title{Dynamic mode locking in a driven colloidal system: experiments and theory}

\author{Michael P.N. Juniper$^1$, Urs Zimmermann$^2$, Arthur V. Straube$^3$, Rut Besseling$^4$, Dirk G.A.L. Aarts$^{1}$, Hartmut L{\"o}wen$^2$, and Roel P.A. Dullens$^{1*}$} 
\affiliation{$^{1}$ Department of Chemistry, Physical and Theoretical Chemistry Laboratory, University of Oxford, South Parks Road, OX1 3QZ Oxford, United Kingdom\\
$^{2}$ Institut f{\"u}r Theoretische Physik II - Soft Matter, Heinrich-Heine-Universit{\"a}t D{\"u}sseldorf, Universit{\"a}tsstra{\ss}e 1, D-40225 D{\"u}sseldorf, Germany\\
$^{3}$ Department of Mathematics and Computer Science, Freie Universit\"at Berlin, Arnimallee 6, 14195 Berlin, Germany\\
$^{4}$ InProcess-LSP, Molenstraat 110, 5342 CC, Oss, Netherlands}

\date{\today}

\begin{abstract}
In this article we examine the dynamics of a colloidal particle driven by a modulated force over a sinusoidal optical potential energy landscape.  Coupling between the competing frequencies of the modulated drive and that of particle motion over the periodic landscape leads to synchronisation of particle motion into discrete modes.  This synchronisation manifests as steps in the average particle velocity, with mode locked steps covering a range of average driving velocities.  The amplitude and frequency dependence of the steps are considered, and compared to results from analytic theory, Langevin Dynamics simulations, and Dynamic Density Functional Theory.  Furthermore, the critical driving velocity is studied, and simulation used to extend the range of conditions accessible in experiments alone.  Finally, state diagrams from experiment, simulation, and theory are used to show the extent of the dynamically locked modes in two dimensions, as a function of both the amplitude and frequency of the modulated drive.

\end{abstract}

\maketitle

\section{Introduction}
\label{mlintro}

Synchronisation is one of the most diverse fundamental physical phenomena \cite{Pikovsky2001}.  From Huygens' pendulum clocks 350 years ago \cite{Birch1756,Bennett2001} to fireflies \cite{Agrawal2013}, applause \cite{Neda2000,Neda2000PRE}, and animals' circadian rhythms \cite{Nicolis2013}, frequency entrainment occurs all over the natural and technological world.  The phenomenon occurs when weakly coupled competing oscillators adjust their rhythms to match each other \cite{Agrawal2013}.  Synchronisation on the micro-scale is of technological importance, as the decreasing size of electronic and mechanical systems demands ever-smaller frequency references \cite{Feng2008}.  Recent developments include electromechanical \cite{Antonio2012,Shim2007} and optomechanical oscillators \cite{Zalalutdinov2003,Zhang2012}, but such systems are limited in their scalability \cite{Zhang2012}.

Dynamic mode locking is a synchronisation phenomenon that occurs when systems with a natural internal frequency are driven by an external modulation.  Competition between the two frequencies leads to coupling, causing the system to synchronise into repeating modes of motion.  Previous work has sought to understand dynamic mode locking in superconductor vortex lattices \cite{Besseling2005,Olsen1998,Kokubo2004,Kokubo2002,Kolton2001}, but the difficulty in visualising such systems makes model systems necessary \cite{Grigorieva1994}.  Other systems showing such resonance behaviour include driven adatoms on atomic surfaces \cite{Aubry1978,Talkner1999}, and Josephson junctions \cite{Kvale1991,Burkov1991,Grimes1968}.  The AC Josephson effect occurs when the tunneling electron pairs at an insulated superconductor junction are driven with an AC and DC current \cite{Josephson1962}.  Regions appear where resistance does not increase with increasing DC current \cite{Josephson1962,Grimes1968}, and the shape of the resulting graph is known as `Shapiro Steps'.  Charge density waves are 
another technologically significant system demonstrating dynamic mode locking, and have been extensively studied experimentally and numerically \cite{Gruner1988,Carpinelli1996,Eichberger2010}.

Model systems composed of colloidal particles in periodic potentials have been studied for a number of years \cite{Sancho2010,Dobnikar2013}, from simple double well potentials \cite{Simon1992,Curran2012,Schmitt2006,Babic2004} to directed motion \cite{Hennig2010,Hasnain2013,Mcdermott2013,Zaidouny2013,Wang2013,TekicMali-book}, particle sorting \cite{Pelton2004}, and kink generation \cite{Bohlein2012} in two-dimensional (2D) optical lattices.  Colloidal systems are easy to manipulate, and have accessible length and time scales, making them attractive models for the study of synchronisation at the micro-scale.  Noise in Brownian systems has been found in theory to induce anomalous diffusion \cite{Lindenberg2005} and stochastic resonance \cite{Shi2014,Reguera2002,Tekic2008}, and rocking-ratchet like potentials have been used in optical and magnetic systems \cite{Arzola2013,Herrera2008}.  The possibility of resonance has also been explored in systems with feedback \cite{Lichtner2012} and random pinning potentials \cite{Chen2007}.  Recent work studied the transport properties of a system of magnetically driven colloidal particles \cite{Straube2013}.  Recent theoretical work also examines the possibility of producing mode locking steps in 2D colloidal monolayers \cite{Song2015,Paronuzzi2016}.

Here, a system of Brownian particles is driven over a sinusoidal optical potential energy landscape by a driving force consisting of constant and modulated parts.  The natural frequency of the particle driven over the optical potential energy landscape by the DC component of the driving force couples to the frequency of the AC component.  As we have shown previously \cite{Juniper2015}, this coupling leads to dynamic mode locking.  This work considers the frequency and amplitude dependence of the synchronisation, through experiments, Langevin Dynamics simulations and Dynamic Density Functional Theory (DDFT).  The three complementary approaches are used together to build a comprehensive picture of dynamic mode locking.  Firstly, the theoretical and simulation approaches are introduced in Section \ref{mltheory}, including an analytical approximation.  The experimental methods are described in Section \ref{mlexp}.  Results from all of the approaches are described and discussed in Section \ref{mlres}, including mode locked steps, state diagrams, and critical driving forces.

\section{Theory and computer simulations}
\label{mltheory}

\subsection{Langevin dynamics}
\label{LangevinDynamics}

To describe a Brownian particle driven by the sum of a constant and a modulated force across a periodic optical potential energy landscape, the overdamped Langevin equation is written as:
\begin{equation}
	\zeta~v(x,t) = \zeta\frac{\textup{d}x}{\textup{d}t} = F_{\textup{DC}} + F_{\textup{mod}}(t) + F_{\textup{T}}(x) + \xi(t),
\end{equation}
where the particle velocity, $v$, at position $x$ and time $t$ depends on the force from the optical potential energy landscape, $F_{\textup{T}}$, the Brownian force, $\xi(t)$ (modelled as Gaussian white noise with a mean of zero and variance of $2\zeta k_{\textup{B}}T$, where $k_{\textup{B}}T$ is thermal energy), the constant driving force, $F_{\textup{DC}}$, the friction coefficient, $\zeta$, and the oscillating driving force,
\begin{equation}
	F_{\textup{mod}}(t)=F_{\textup{AC}}\cos\left(\omega t\right).
\end{equation}
Here, $F_{\textup{AC}}$ is the amplitude of the modulated driving force, and $\omega=2\uppi\nu$ is the angular frequency, where $\nu$ is the frequency of the applied oscillation.  Note that in this paper, `DC' and `AC' are used only in analogy to direct- and alternating-current, and refer to constant- and oscillating-velocity drives respectively.

The optical potential energy landscape, $U_{\textup{T}}(x)$, is taken to be sinusoidal, as described in references \cite{Juniper2012, Juniper2015, Juniper2016}:
\begin{align}\label{ut}
	U_{\textup{T}}(x)= -\frac{2\sqrt{2\uppi}V_0^{3/2}}{\lambda k^{1/2}}\left[\frac{1}{2}+\exp\left(-\frac{2\uppi ^2V_0}{\lambda^2 k}\right)\cos\left(\frac{2\uppi x}{\lambda}\right)\right],
\end{align}
where $k$ is the trap stiffness, $V_0$ is the trap strength, and $\lambda$ is the wavelength of the landscape.  Equation \ref{ut} leads to an optical force \cite{Juniper2016}:
\begin{equation}
	F_{\textup{T}}=-\frac{\partial U_{\textup{T}}}{\partial x}=- F_{\textup{C}}\sin\left(\frac{2\uppi x}{\lambda}\right),
\end{equation}
where the critical driving force, $F_{\textup{C}}$, is given by the following equation \cite{Juniper2016}:
\begin{equation}\label{critdriv}
	F_{\textup{C}}=\frac{4\sqrt{2}(\uppi V_0)^{3/2}}{\lambda^2 k^{1/2}}\exp\left(-\frac{2\uppi ^2V_0}{\lambda^2 k}\right).
\end{equation}
Thus, the full equation of motion for a particle driven by DC and AC driving forces over a sinusoidal optical potential energy landscape is given by:
\begin{align}\label{acdceom}
	\zeta~v(x,t) &= \zeta~\frac{\textup{d}x(t)}{\textup{d}t} \nonumber\\
	&= F_{\textup{DC}} + F_{\textup{AC}}\cos\left(2\uppi\nu t\right) - F_{\textup{C}}\sin\left(\frac{2\uppi x}{\lambda}\right) + \xi(t).
\end{align}
Note that the critical driving force, $F_{\textup{C}}$, is a property of the landscape, and is equal to the DC critical driving force found in reference \cite{Juniper2016}.  As the total driving force in equation \ref{acdceom} is a sum of the DC and time dependent AC contributions, the critical DC driving force required to overcome pinning depends on the amplitude and frequency of the modulated component of the driving force.

\subsection{The `high frequency' theory}
\label{highfreqlimit}

\begin{figure*}[t]
	\centering
	\includegraphics[width=1\textwidth]{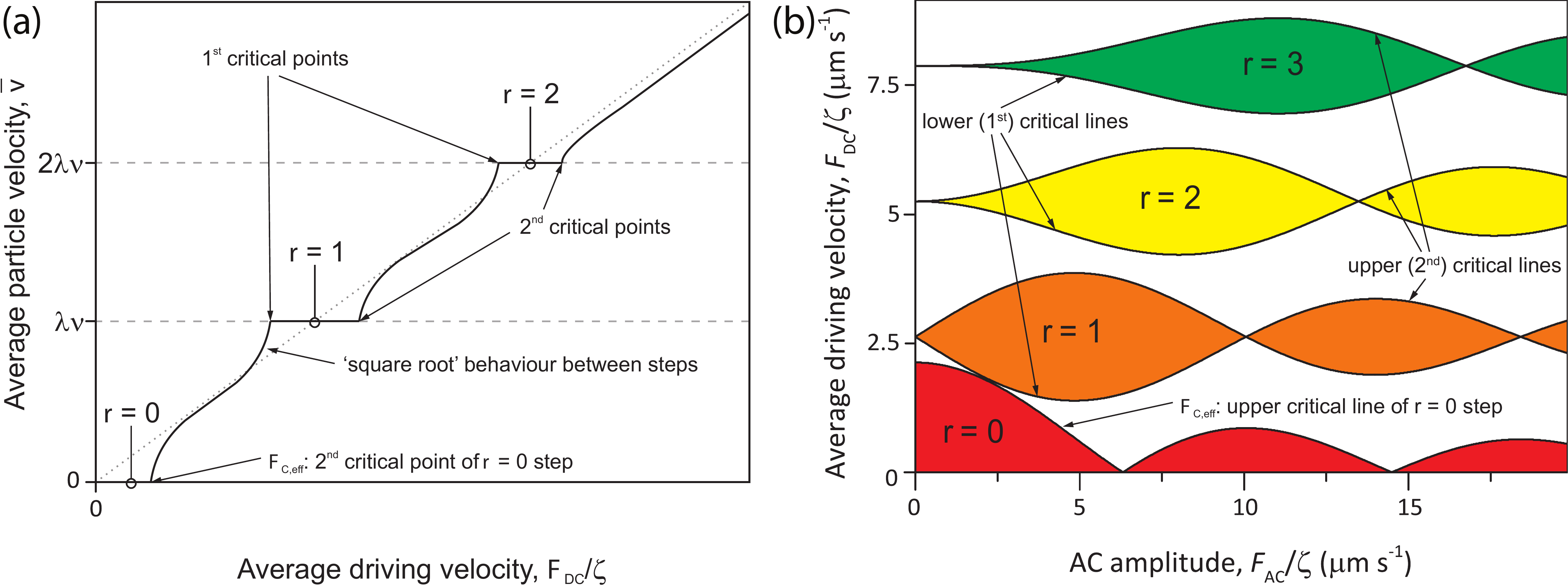}
	\caption{\small{Mode locked steps and state diagram.  (a) Schematic of average particle velocity as a function of average driving velocity (see equation \ref{mlhighfreq}; not to scale).  Mode locked steps lie at $r\lambda\nu$, between pairs of critical points.  (b) State diagram calculated from equation \ref{Fchighw}, for a periodic optical potential energy landscape with a trap spacing of $\lambda=3.5~\upmu$m, and a frequency of $\nu=\frac{3}{4}$ Hz.  Critical points defining a step in (a) become critical lines enclosing mode locked regions in the state diagram in (b).}}
	\label{ml_theory_sd}
\end{figure*}

While the equation of motion in equation \ref{acdceom} is not analytically soluble, useful insight can be obtained in the limit of high driving frequency ($\nu\gg F_{\textup{C}}/\lambda\zeta$) in the absence of noise.  Within this approximation, it is possible to obtain an effective Adler equation \cite{Adler1946, Goldstein2009} similar to that found for the case of constant drive alone \cite{Juniper2016}.  Thus an expression for the average velocity may be written (see Appendix A for full details):

\begin{equation}\label{mlhighfreq}
	 \overline{v} = \left\{
\begin{aligned}&	r\lambda\nu, && \mbox{if }\quad |\Delta F_{\textup{DC}}| < \left|F_{\textup{C}} J_{-r}\left(\frac{F_{\textup{AC}}}{\lambda\nu\zeta}\right)\right|;\\
	& r\lambda\nu \pm \frac{1}{\zeta}\sqrt{\Delta F_{\textup{DC}}^2-F_{\textup{C}}^2 J_{-r}^2\left(\frac{F_{\textup{AC}}}{\lambda\nu\zeta}\right)}, && \mbox{if }\quad |\Delta F_{\textup{DC}}| > \left|F_{\textup{C}} J_{-r}\left(\frac{F_{\textup{AC}}}{\lambda\nu\zeta}\right)\right|, \quad \Delta F_{\textup{DC}} \gtrless 0,\end{aligned} \right.
\end{equation}
where $\Delta F_{\textup{DC}}=F_{\textup{DC}}-r \lambda \nu\zeta$, $r=0,\pm 1,\pm 2,\ldots$, and $J_{m}$ is the $m$th order Bessel function of the first kind.  This `square root law' expression is the AC driven counterpart to the simpler form found for the case of DC drive alone \cite{Juniper2016}:
\begin{equation*}\label{fdc2fc2}
	 \overline{v} = \left\{
\begin{aligned}& 0, & \mbox{if } F_{\textup{DC}} \leq F_{\textup{C}};\\
	& \frac{1}{\zeta}\sqrt{F_{\textup{DC}}^2-F_{\textup{C}}^2}, & \mbox{if } F_{\textup{DC}} > F_{\textup{C}}. \end{aligned}\right.
\end{equation*}
The condition for the approximation, $\nu\gg F_{\textup{C}}/\lambda\zeta$ (see equation \ref{condition}, Appendix A), means that on a landscape with a trap spacing of $\lambda=3.5~\upmu$m ($F_{\textup{C}}/\zeta\approx 1.8~\upmu$m s$^{-1}$), the high frequency regime is valid when $\nu\gg 0.5$ Hz.

The dependence of the average particle velocity, $\overline{v}$, on the driving velocity, $F_{\textup{DC}}/\zeta$, according to equation \ref{mlhighfreq} is shown schematically in figure \ref{ml_theory_sd}(a).  Equation \ref{mlhighfreq} describes mean particle velocity `above' and `below' critical points, with two critical points found for every absolute value of $r$, in contrast to the DC only case, which has only a single $F_{\textup{C}}$.  Between each pair of critical points, a `subcritical' regime exists, where the particle velocity is constant, corresponding to mode locked steps.  The form of this dependence is analogous to the Shapiro Steps seen in Josephson Junctions \cite{Grimes1968}, and also in charge density wave systems \cite{Gruner1988} and vortex lattices \cite{Kokubo2002, Kokubo2004, Kolton2001, Reichhardt2000}.

The two critical points $F_{\textup{CRIT}}$ at the ends of resonant step $r$ are found by determining $F_{\textup{DC}}$ at the condition where the two different solutions in equation \ref{mlhighfreq} coincide, $\Delta F_{\textup{DC}}=\pm |F_{\textup{C}}J_{-r}(F_{\textup{AC}}/(\lambda\nu\zeta))|$, at which point the square root vanishes.  As a result, by recalling the definition of $\Delta F_{\textup{DC}}$ and by replacing $F_{\textup{DC}}$ with $F_{\textup{CRIT}}$, the following is obtained:
\begin{equation}\label{Fchighw}
	F_{\textup{CRIT}}/\zeta = r\lambda\nu \pm \frac{F_{\textup{C}}}{\zeta}\left| J_{-r}\left(\frac{F_{\textup{AC}}}{\lambda\nu\zeta}\right)\right|.
\end{equation}
The amplitude ($F_{\textup{AC}}$) and frequency ($\nu$) dependence of $F_{\textup{CRIT}}$ defines state diagrams, with regions containing locked modes enclosed by pairs of critical lines.  Figure \ref{ml_theory_sd}(b) shows such a state diagram as a function of $F_{\textup{AC}}$, for a particle driven with a frequency of $\nu=\frac{3}{4}$~Hz across an optical landscape with a trap spacing of $\lambda=3.5~\upmu$m.  Each colour and value of $r$ represents a single mode locked velocity.  The state diagram is formed from twisted `Arnold Tongues' \cite{Pikovsky2001}, where each separated region of the same colour actually represents a different dynamic mode with the same average velocity \cite{Juniper2015}.  The second critical point of the `zeroth' step appears as the effective critical driving velocity, $F_{\textup{C,eff}}$, below which the particle is pinned to the landscape and does not slide.

\subsection{Dynamic Density Functional Theory}
\label{ddft}

The Langevin picture is stochastically equivalent to the Smoluchowski picture, in which the temporal evolution of the probability density distribution, $p(x, t)$, of the particle position is studied rather than the stochastic trajectories of individual particles. The Smoluchowski equation can be seen as a special case of the Dynamical Density Functional Theory (DDFT) in the absence of interparticle interactions \cite{Tarazona_Marconi_JCP,Archer_Evans_JCP,Espanol_Lowen_JCP_2009}.  The governing equation for the probability density distribution is given by
\begin{align}\label{Smo_eq2}
	\frac{\partial p(x,t)}{\partial t} &= D \frac{\partial^2 p(x,t)}{\partial x^2} + \frac{1}{\zeta}\frac{\partial}{\partial x} \left(F(x,t) p(x,t) \right),
\end{align}
where $D$ is the diffusion coefficient and $F(x,t) = F_\text{DC} + F_\text{AC}\cos(2\uppi\nu t) - F_\text{C} \sin(2\uppi x/\lambda)$ is the total force acting on the particle. Equation \ref{Smo_eq2} is solved numerically using a finite volume partial differential equation solver \cite{FiPy:2009}.  As an initial condition $p(x, t=0)$, a very narrow Gaussian distribution is chosen.  See Appendix B for more details.

Within the Smoluchowski picture, averages of statistical quantities are defined by weighting these quantities with the particle probability distribution $p(x,t)$, i.e. $\langle a \rangle(t) = \int_{-\infty}^\infty \! \mathrm{d}x \; a(x) p(x,t)$.  These averages are stochastically equivalent to noise averages performed in the Langevin picture.  Thus, the mean particle position is $\langle x\rangle(t)$.  The mean velocity is further defined as the change in the mean particle position in time:
\begin{equation}
	\overline{v} = \overline{\frac{\mathrm{d} \langle x \rangle}{\mathrm{d}t}},
\end{equation}
where overbar denotes a time average.  As a measure of the fluctuations around the mean particle trajectory, the variance of the particle probability distribution is considered:
\begin{equation}
	\sigma^2(t) = \langle [x - \langle x \rangle]^2 \rangle(t).
\end{equation}
In the context of this work, if the standard deviation, $\sigma(t)$, is much smaller than the trap spacing then almost all possible particle trajectories end up in the same trap as the mean particle position after time $t$.  If the standard deviation is larger than $\lambda$ then possible particle trajectories end up distributed in potential wells surrounding the mean.  Particle fluctuations around the mean position may be quantified using an effective long--time diffusion coefficient, defined from the variance:
\begin{equation}\label{Deff}
	D_\text{eff} = \lim_{t \rightarrow \infty} \frac{\sigma^2(t)}{2t}.
\end{equation}

\section{Experimental methods}
\label{mlexp}

\subsection{Colloidal model system}

The colloidal system is composed of Dynabeads M-270 carboxylic acid (diameter $3~\upmu$m), in 20\% EtOH$_{\textup{aq}}$, held in a quartz glass sample cell (Hellma) with internal dimensions of 9 $\times$ 20 $\times$ 0.2 mm.  Particles are much more dense than the solvent, and sediment into a single layer near the bottom of the sample cell.  The coefficient of friction, $\zeta$, is found from diffusion to be $9.19 \times 10^{-8}$~kg~s$^{-1}$, slightly higher than expected from Stokes friction ($\zeta_{Stokes} = 6\uppi\eta a$ with $\eta$ the viscosity), due to the proximity of the particles to the wall.  Particle concentration is low, so that only a single particle is visible in the field of view.

\subsection{Experimental setup and parameters}

The experimental setup consists of an infra-red (1064~nm) laser, controlled using a pair of perpendicular acousto-optical deflectors, and focused using a 50$\times$, NA=0.55 microscope objective \cite{Juniper2012}.  The one-dimensional periodic optical landscape, with trap spacing $\lambda=3.5~\upmu$m, is generated in Aresis Tweez software controlled from a LabView interface.  A landscape with this trap spacing may be treated as sinusoidal, as shown in reference \cite{Juniper2016}.  The traps are time-shared at 5 kHz, such that on the time scale of the particles (with a Brownian time of $\sim50$~s, and at least $\sim\frac{1}{3}$~s to be driven one trap spacing at a given $F_{\textup{DC}}$), the traps form a constant potential energy landscape.  The laser power and the total number of traps are held constant throughout the experiments, so that the laser power per trap is consistent.  A laser power of 350 mW is set and 46 traps are used, corresponding to $\sim 0.75$ mW per trap at the sample.  This gives typical values of trap stiffness, $k=3.8 \times 10^{-7}$ kg s$^{-2}$, and trap strength, $V_0~=~90~ k_{\textup{B}}T$ \cite{Juniper2012, Juniper2016}.

The driving force is provided by a PI-542.2CD piezo-stage, controlled using the LabView interface, at driving velocities of $0.05~\le~F_{\textup{DC}}/\zeta~\le~8~\upmu$m~s$^{-1}$.  AC driving velocity is added to the DC drive, with an amplitude $0.4~\le~F_{\textup{AC}}/\zeta~\le~14~\upmu$m~s$^{-1}$, and a frequency of $\frac{1}{10}$~Hz~$\le \nu \le 2$ Hz.

Images are focused onto a Ximea CMOS camera using a 40$\times$, NA$=0.50$ microscope objective, and the particle position is recorded live at 40 Hz from the camera image.

\subsection{Average velocity experiments}
\label{findingPPs}

To obtain plots of average particle velocity against driving velocity, six repeats across the whole potential landscape are made at each driving velocity for each amplitude and frequency of the AC drive.    Average velocity, $\overline{v}$, is found by linearly fitting the particle trajectory, $x(t)$, over an integer number of periods of the oscillation.

\subsection{Critical driving velocity experiments}
\label{effcritvel}

The critical DC driving velocity is defined as the DC driving velocity at which the particle starts to slide irreversibly across the optical potential energy landscape.  It is found by iterating the DC driving velocity, with a maximum resolution of $0.05~\upmu$m s$^{-1}$.  A particle is said to be pinned if it still returns to its starting lattice position after the stage has moved $100~\upmu$m, or three minutes has elapsed, whichever happens first.  The region in which the particle does not irreversibly slide is essentially the zeroth mode locked step, and the effective critical driving velocity is therefore the second critical point of this step (see section \ref{highfreqlimit}).

The critical driving velocity found here is \emph{not} the critical driving velocity found in our previous work (\cite{Juniper2016}), $F_{\textup{C}}/\zeta$, because the total driving force in equation \ref{acdceom} is a sum of the DC and time dependent AC contributions.  Therefore the critical driving velocity measured here is an \emph{effective} critical driving velocity, $F_{\textup{C,eff}}/\zeta$, as it is not purely a property of the landscape.  However, clearly when $F_{\textup{AC}}=0$, $F_{\textup{C,eff}}\equiv F_{\textup{C}}$.

\begin{figure*}[t]
	\includegraphics[width=1.0\textwidth]{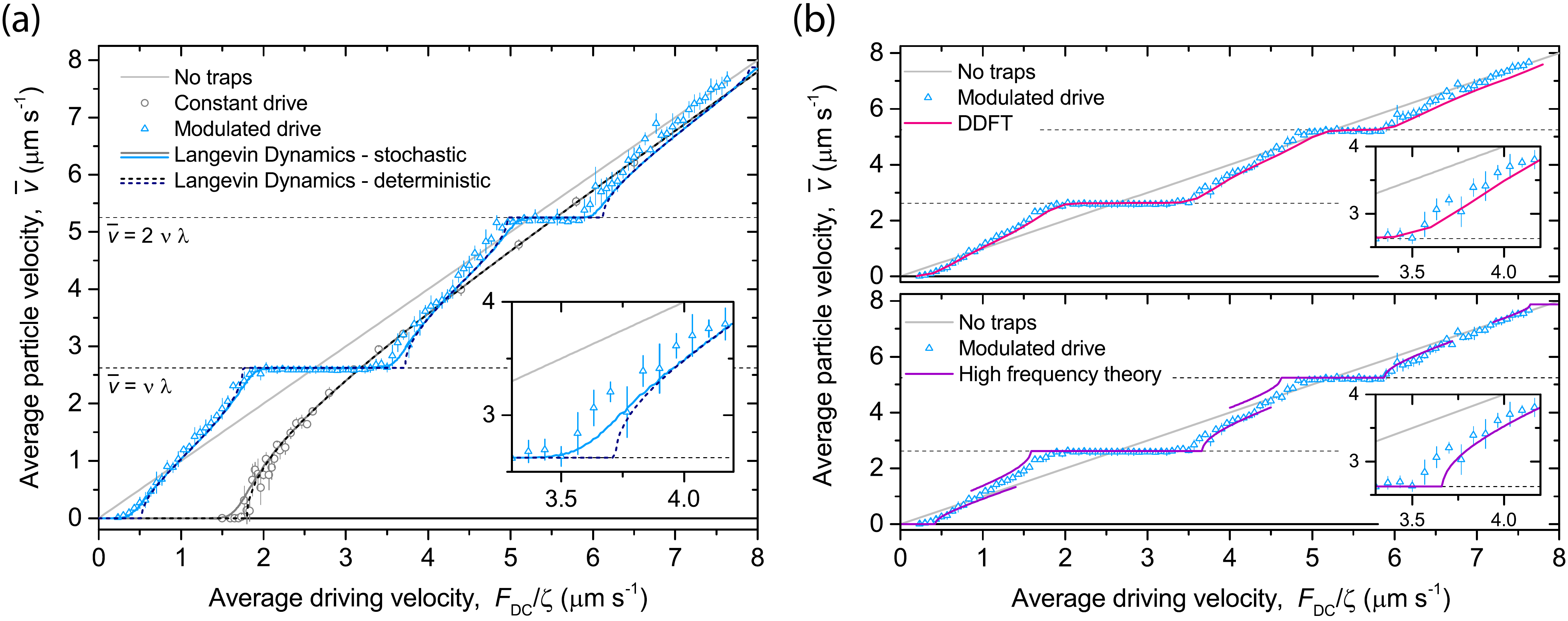}
	\centering
	\caption{\small{Average particle velocity, $\overline{v}$, against average driving velocity, $F_{\textup{DC}}$, for both constant and modulated drives, from experiments, simulations, high frequency theory, and Dynamic Density Functional Theory (DDFT).  (a) Comparison of experimental data and Langevin Dynamics simulations.  \textcolor{Grey}{$\medcircle$} constant drive: $F_{\textup{AC}}=0$; \textcolor{Cerulean}{$\medtriangleup$} modulated drive: $F_{\textup{AC}}/\zeta=5.2~\upmu$m s$^{-1}$, $\nu=\frac{3}{4}$ Hz (error bars represent the standard deviation of the repeats); solid lines: Langevin Dynamics results; dashed lines: Langevin Dynamics results with no noise term; grey line: no traps calibration; horizontal dashed lines: step positions $\overline{v}=n\lambda\nu$.  Includes experimental data from references \cite{Juniper2015} and \cite{Juniper2016}.  (b) Comparison of experimental data with DDFT results (upper panel), and high frequency theory (lower panel).  Insets highlight the second critical point of the first step.}}
	\label{firstmlgraph}
\end{figure*}

\section{Results and discussion}
\label{mlres}

Results are presented which show the amplitude and frequency dependence of the mode locked steps and state diagrams illustrated in figure \ref{ml_theory_sd}.  Experimental results and Dynamic Density Functional Theory (DDFT) computations are compared to Langevin Dynamics (LD) simulations and the analytic approximation for the high frequency limit (Section \ref{highfreqlimit}) as appropriate.  In general, good quantitative agreement is found between the various approaches.

\begin{figure}[t]
	\includegraphics[width=1\columnwidth]{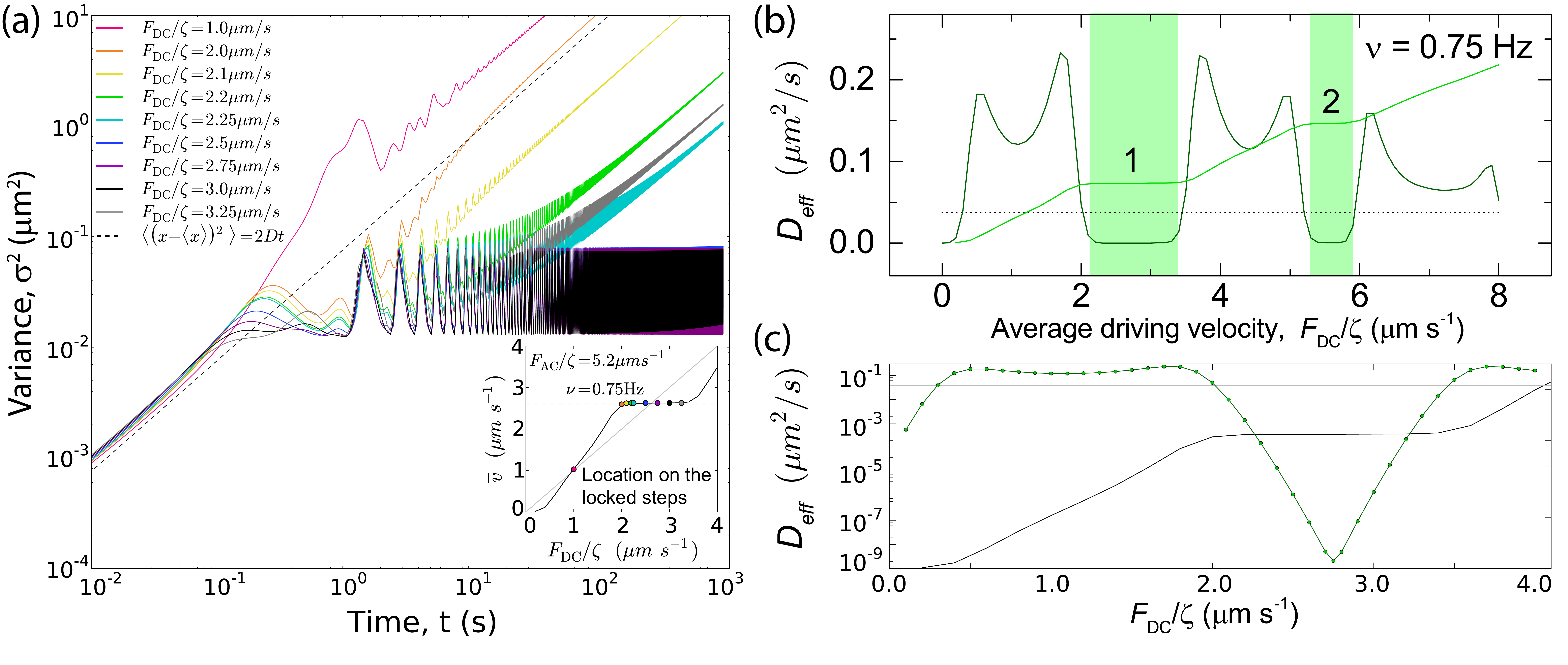}
	\centering
	\caption{\small{Variance and long-time diffusion from DDFT calculations, for $\nu~=~\frac{3}{4}$~Hz~and~$F_{\textup{AC}}/\zeta~=~5.2~\upmu$m~s$^{-1}$.  (a) Double logarithmic plot of variance versus time for various drives $F_\text{DC}$. The black dashed line refers to a Brownian particle in the absence of an external potential. The inset shows the location of the respective state points on the locked steps for the lines in the main panel.  (b) Effective long--time diffusion coefficient as a function of average driving velocity.  Dark line shows $D_{\textup{eff}}$, pale line shows the mode locked staircase calculated from DDFT for comparison, and does not match the $y$-axis scale.  Step extents are highlighted by the coloured bands, and the dotted line indicates the diffusion coefficient of a free particle.  (c) $D_{\textup{eff}}$ for the first step, plotted on a log scale.}}
	\label{centvarfig}
\end{figure}

\subsection{The mode locking steps}
\label{devils}

The effect of introducing the oscillating force term to the equation of motion (equation \ref{acdceom}) on $\overline{v}$ as a function of $F_{\textup{DC}}/\zeta$ is shown in figure \ref{firstmlgraph}(a).  Here, data with a modulated force of amplitude $F_{\textup{AC}}/\zeta=5.2~\upmu$m s$^{-1}$ and frequency $\nu=\frac{3}{4}$ Hz (\textcolor{Cerulean}{$\medtriangleup$}) is compared to the $F_{\textup{AC}}=0$ (\textcolor{Grey}{$\medcircle$}, i.e. DC drive only) case for a landscape of trap spacing $\lambda=3.5~\upmu$m (see \cite{Juniper2015,Juniper2016}).  The case of $F_{\textup{AC}}\ne 0$ follows the `Shapiro Steps' form illustrated in figure \ref{ml_theory_sd}(a).  The effective critical driving velocity for the modulated case is almost zero, after which $\overline{v}$ increases until it is significantly larger than that expected for a free particle (grey line), implying that the particle is moving on average more quickly than the piezo stage.  The average velocity then plateaus on the first resonant step, at $\overline{v}=1\lambda\nu = 2.625~\upmu$m s$^{-1}$.  The step extends over a range of driving velocities, and then $\overline{v}$ increases after the second critical point, to meet another step, at twice the average particle velocity of the first.

The solid and dashed lines on figure \ref{firstmlgraph}(a) show results from Langevin Dynamics simulations, both with and without the noise term.  The steps found from the experiments are faithfully reproduced by the simulations, with the inclusion of noise obviously important in this system of Brownian particles.  The effect of noise is important in the vicinity of the critical points, where it is seen to round the edges of the steps.  Figure \ref{firstmlgraph}(b) compares results from DDFT (upper panel), and the high frequency approximation (lower panel, see section \ref{highfreqlimit}) to the experimental data.  The high frequency theory results, from equation \ref{mlhighfreq}, are calculated with respect to each critical point, and it is notable that although the step positions are largely captured, the results between steps, from adjacent critical points, do not necessarily agree.

\subsubsection*{Variance and diffusion}

Next, we perform DDFT calculations and consider fluctuations around the mean particle position, which strongly depend on whether or not the system is mode locked.  In figure \ref{centvarfig}(a) the variance is shown as a function of time for the conditions considered above ($F_{\textup{AC}}/\zeta=5.2~\upmu$m s$^{-1}$, $\nu=\frac{3}{4}$~Hz). The displayed numerical data correspond to states on the mode locked steps around the first step (see inset).  For unlocked states ($1.0~\le~F_{\textup{DC}}/\zeta~\le~2.0~\upmu$m~s$^{-1}$) the variance grows rapidly, corresponding to an effective diffusion much larger than the free diffusion (dashed line).  In the mode locked states ($2.5~\le~F_{\textup{DC}}/\zeta~\le~3.25~\upmu$m~s$^{-1}$) the variance reaches a long--lived plateau where the diffusion is (nearly) zero, before eventually crossing over.  Similar intermediate plateaus have also been observed in underdamped systems \cite{Guo2014, Marchenko2012, Saikia2009} and static systems ($F_{\textup{AC}}~=~0$) \cite{Lindenberg2007, Emary2012}.

The effective long-time diffusion coefficient (equation \ref{Deff}) is the limit of $\sigma^2(t)/2t$ as $t \rightarrow \infty$.  Plotting $D_{\textup{eff}}$ as a function of average driving velocity offers additional insight into the mode locked steps.  Figure \ref{centvarfig}(b) shows that effective diffusion is close to zero at the mode locked steps, and much higher between.  The typical double-peak signature for $D_{\textup{eff}}$, as described in \cite{Reguera2002}, is recovered.  The low $D_{\textup{eff}}$ values in the locked regions result from a vanishing influence of thermal noise which can also be found in related systems \cite{Wiesenfeld1987, Crommie1991}.  This is a symptom of the predictability of the locked state: when the particle is locked into a particular mode of motion, its position on the periodic landscape after a certain time depends purely on the driving conditions.  $D_{\textup{eff}}$ is highest in unlocked states on the cusp of synchronisation conditions, as a small perturbation may cause the particle to jump to the next potential well, or stay in the present one.  This corresponds to the discontinuities at the critical points in the schematic in figure \ref{ml_theory_sd}(a).  Figure \ref{centvarfig}(c) shows $D_{\textup{eff}}$ for the first step on a log scale, showing that it decreases by $\sim8$ orders of magnitude between the unlocked and locked states.  To put this into context, the lowest effective long--time diffusion coefficient $D_{\textup{eff}}\approx2.1\times10^{-9}~\upmu$m$^2$~s$^{-1}$ corresponds to the particle being one lattice spacing away from the predicted position after approximately 45 years.

\begin{figure}[t!]
	\centering	
	\includegraphics[width=1\columnwidth]{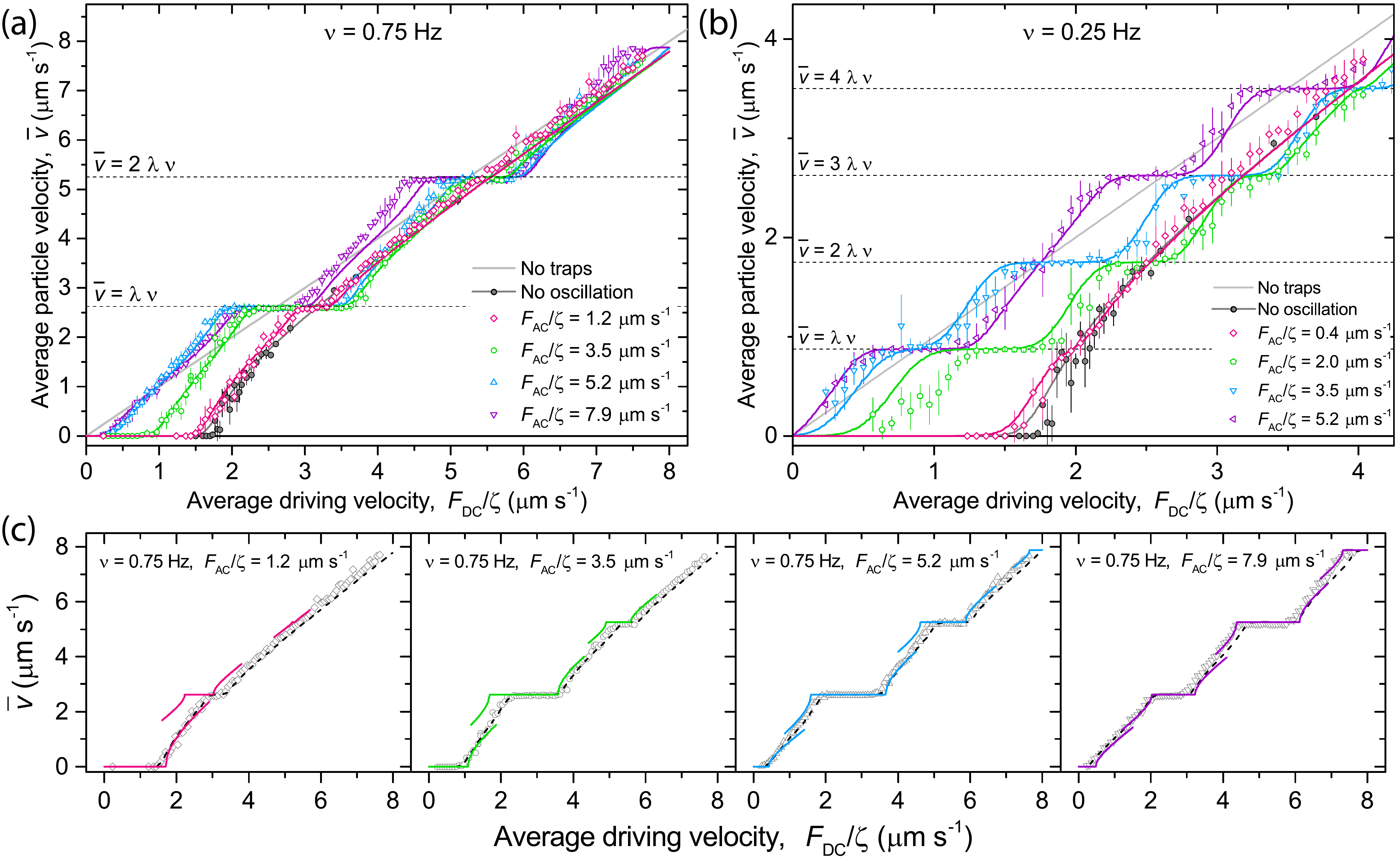}
	\caption{\small{Average particle velocity as a function of average driving velocity, at constant AC frequency of (a) $\frac{3}{4}$ Hz and (b)~$\frac{1}{4}$~Hz.  Symbols show experimental data, solid lines show results of Langevin Dynamics simulations.  Dashed lines indicate step positions $n \lambda \nu$.  Dark grey circles and line indicate case for no oscillation \cite{Juniper2016}, light grey line indicates case for no traps.  (c) Results for high frequency theory (solid lines), for each of the cases in panel (a) for $\nu=\frac{3}{4}$~Hz, compared to LD simulations (dashed lines) and experiments (symbols).}}
	\label{allw075}
\end{figure}

\subsection{Dependence on the amplitude}
As the synchronisation condition depends only on the trap spacing and the modulation frequency, changing the modulation amplitude alone does not alter the step velocities.  Figure \ref{allw075}(a) shows mode locking steps obtained from both experiment and Langevin Dynamics simulations for four different amplitudes, at a frequency of $\nu=\frac{3}{4}$ Hz.  This shows that there is, however, a strong dependence of the width of the locked step on the oscillation amplitude.  For very low amplitude there is a small visible first step, giving a deviation from the zero oscillation data, but no second step is observed; the points lie on top of the $F_{\textup{AC}}=0$ line.  As amplitude increases, the first step increases in width, and a second step appears and widens.  The first step then appears to narrow.  There is generally a good agreement between the experimental data and the LD simulations, with small deviations possibly due to experimental uncertainties such as the variability of the laser power during the experiment.

Figure \ref{allw075}(b) shows data at a lower frequency of $\nu=\frac{1}{4}$ Hz, for four amplitudes.  The lower frequency means that a larger number of steps appear in the same range of particle velocities.  Four steps are visible in the range shown (which is smaller than that in figure \ref{allw075}(a)), with step width varying widely.  Notably, the $F_{\textup{AC}}/\zeta=5.2~\upmu$m~s$^{-1}$ line has three steps at $n=1,3$ and $4$, but no step is visible at $n=2$.

It is pertinent at this juncture to compare the results for $\nu=\frac{3}{4}$~Hz to the high frequency theory (equation \ref{mlhighfreq}).  Figure \ref{allw075}(c) shows each of the four sets of conditions in panel (a), with the results for each critical point from equation \ref{mlhighfreq} (solid lines) compared to the LD results (dashed lines) and experimental results (symbols) from panel (a).  The first observation is that the high frequency approximation appears to better match the data at higher amplitudes.  The most likely reason for this is that the Bessel function in equation \ref{mlhighfreq} gets smaller as the argument, which is proportional to $F_{\textup{AC}}$, increases (see equation \ref{condition} in Appendix A).  This means that at a given frequency the condition setting the validity of the high frequency approximation is fulfilled better for higher $F_{\textup{AC}}$.  Note also that the theory consistently overestimates the step width, as it is deterministic, whereas the critical points in experiments and simulations are somewhat rounded by noise.  Finally, it may be seen that as in figure \ref{firstmlgraph}(b), the lines between steps determined from different critical points do not overlap, as the `square-root law' sections are only valid in the close vicinity of their critical points.

\subsubsection*{State diagram: Low frequency regime}
\label{lowfreqstatediag}

\begin{figure*}[t]
	\includegraphics[width=1.0\textwidth]{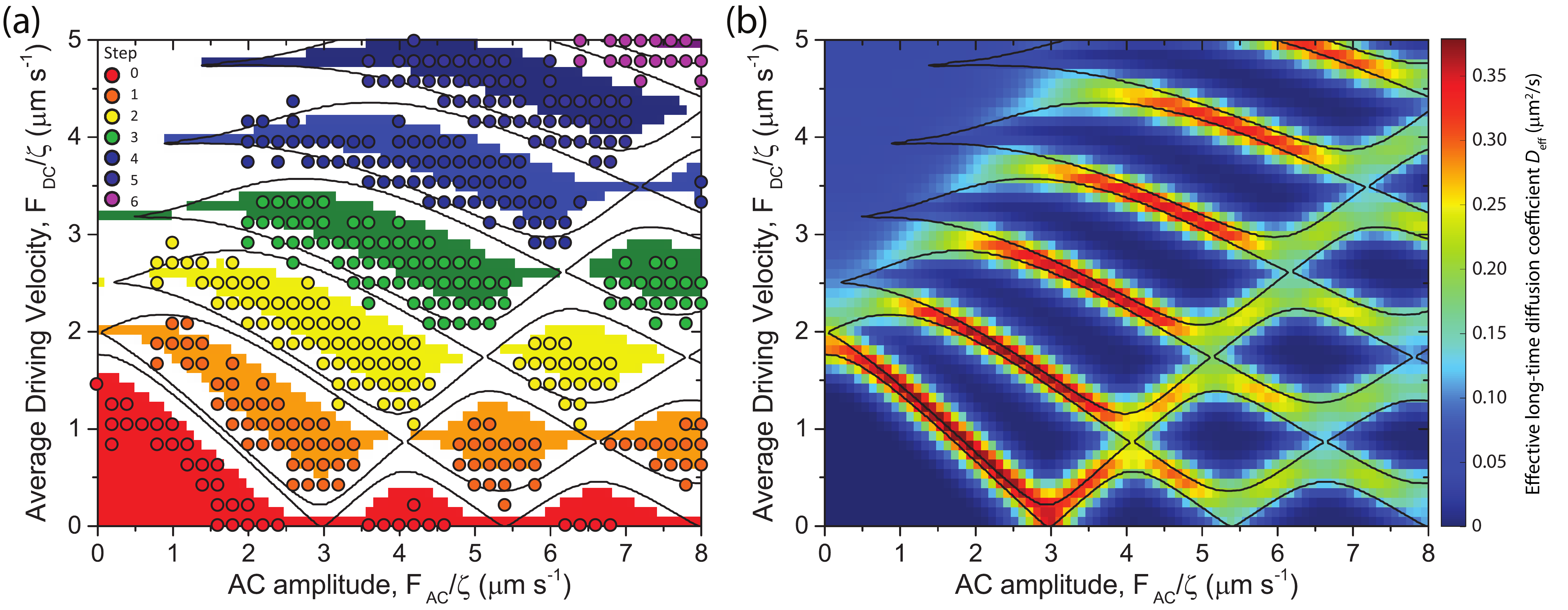}
	\centering
	\caption{\small{State diagram showing locked dynamic modes at a driving frequency of $\frac{1}{4}$ Hz.  (a) Coloured dots: locked states found in experiments (data from reference \cite{Juniper2015}).  Step numbers, $n$, are the net number of steps taken by the particle: regions of the same colour represent the same net forward particle motion.  Coloured regions in the background represent locked states determined from DDFT; white regions are unlocked.  Solid lines show critical lines calculated from LD simulations in the absence of noise.  (b) Colour scale represents the effective diffusion coefficient, $D_{\textup{eff}}$, calculated from DDFT, solid lines are critical lines found from LD simulations as in (a).}}
	\label{state_diag_025}
\end{figure*}

\begin{figure*}[t]
	\includegraphics[width=1.0\textwidth]{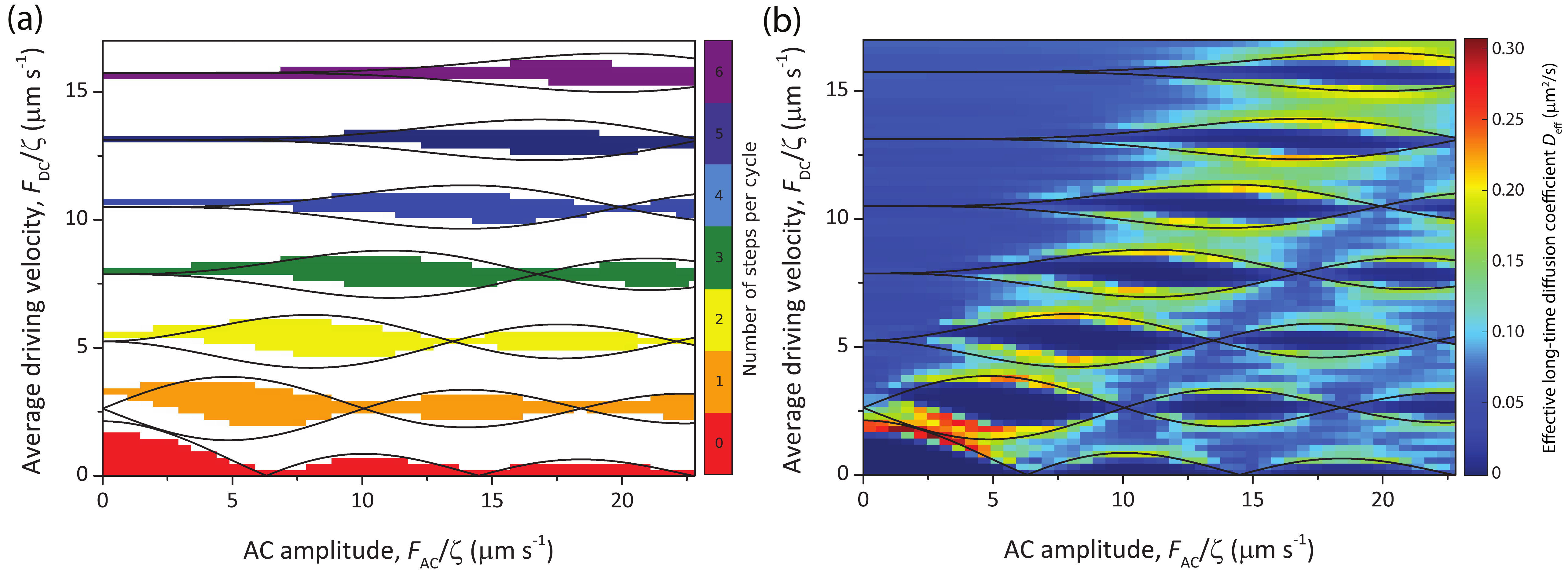}
	\centering
	\caption{\small{State diagram showing all locked dynamic modes at a driving frequency of $\frac{3}{4}$ Hz, over a range of AC amplitudes and average driving velocities.  (a) Coloured regions represent locked states as determined from DDFT, white regions are unlocked.  Locked states are numbered by an integer $n$, the net number of steps taken by the particle.  Solid lines show critical lines calculated from equation \ref{Fchighw}, and represent theoretical state boundaries.  (b) Colour scale represents the effective diffusion coefficient, $D_{\textup{eff}}$, calculated from DDFT, solid lines are again calculated from equation \ref{Fchighw}.  Locked states appear as dark blue.}}
	\label{state_diag_075}
\end{figure*}

As was shown in figure \ref{ml_theory_sd}(b), state diagrams may be constructed which show the extent of dynamic mode locking as a function of $F_{\textup{AC}}/\zeta$ and $F_{\textup{DC}}/\zeta$.  In our previous work \cite{Juniper2015}, we used such a plot to locate numerous dynamic modes for a driving frequency of $\frac{1}{4}$~Hz.  In figure \ref{state_diag_025}(a) we compare these prior experimental results (circles coloured according to the integer step number, $n$) with locked regions from DDFT (background colour) and critical lines from LD simulations in the absence of noise.  Note that points corresponding to unlocked states are not shown, for ease of interpretation.  There is very good agreement between the DDFT and the LD results, with the only difference being that the regions calculated from DDFT are smaller, due to the presence of noise.  Both are in good agreement with the experimental data, except that locked states appear to be found at a slightly higher range of driving velocities in the simulations.

Figure \ref{state_diag_025}(b) again shows critical lines from LD simulations, plotted over the effective diffusion coefficient, $D_{\textup{eff}}$, obtained from DDFT.  Locked states appear as dark blue regions, with fairly broad boundaries where step edges are smoothed by noise.  Between the locked states the effective diffusion is higher, as was seen in figure \ref{centvarfig}(b).  $D_{\textup{eff}}$ is particularly high between closely spaced locked modes, as a small perturbation may lead to the particle becoming temporarily trapped in one mode or the other.  The related enhanced increase in variance manifests in the experimental data as wider error bars between the modes in figure \ref{allw075}(b).  That the diffusion is so high in these regions probably contributes to the mismatch between the experimental and simulation data in panel (a) - a small change in the experimental conditions could cause the particle to cross mode boundaries.

It is nice to observe in figure \ref{state_diag_025} that the mode boundary lines oscillate, with the first and second critical lines crossing and swapping identity between the modes.  The upshot of this oscillation is that there are conditions in which certain locked modes do not appear, for example it is clear that $F_{\textup{AC}}/\zeta=5.2~\upmu$m~s$^{-1}$ lies between two regions with $n=2$, which corresponds exactly to the missing step observed in figure \ref{allw075}(b).  This effect is of course mirrored in the critical driving velocity line, being the upper mode boundary of the zeroth step, resulting in some conditions where the critical driving velocity is zero, for example $F_{\textup{AC}}/\zeta=5.2~\upmu$m~s$^{-1}$ again.

\subsubsection*{State diagram: High frequency regime}
\label{highfreqstatediag}

In section \ref{highfreqlimit}, an analytical expression (equation \ref{Fchighw}) was obtained which could predict the locations of the first and second critical points for each mode locked step.  It was found that this expression should be valid in the region where $\nu \gg 0.5$ Hz.  It is not possible to probe this region in detail in the experiments, as at higher frequencies, increased particle velocities are required to obtain higher modes, with a resulting loss in resolution.  However, using DDFT it is possible to examine this regime, and obtain a state diagram similar to that from experiment.  Figure \ref{state_diag_075}(a) shows a state diagram calculated via DDFT for $\nu=\frac{3}{4}$~Hz, with the mode locked regions being represented by colours, as in figure \ref{state_diag_025}(a).  Also plotted on figure \ref{state_diag_075}(a) are lines calculated from equation \ref{Fchighw}.  There is a remarkably good agreement between the DDFT results and the analytical prediction, showing that the approximation is valid to surprisingly low frequencies.  The main deviation occurs at lower amplitudes, as was seen in figure \ref{allw075}(c). As as in figure \ref{state_diag_025}(a), the locked regions from DDFT are slightly smaller than the space between the critical lines, due to the noise term which must necessarily be ommited from the theory.

\begin{figure}[t]
	\centering
	\includegraphics[width=.5\textwidth]{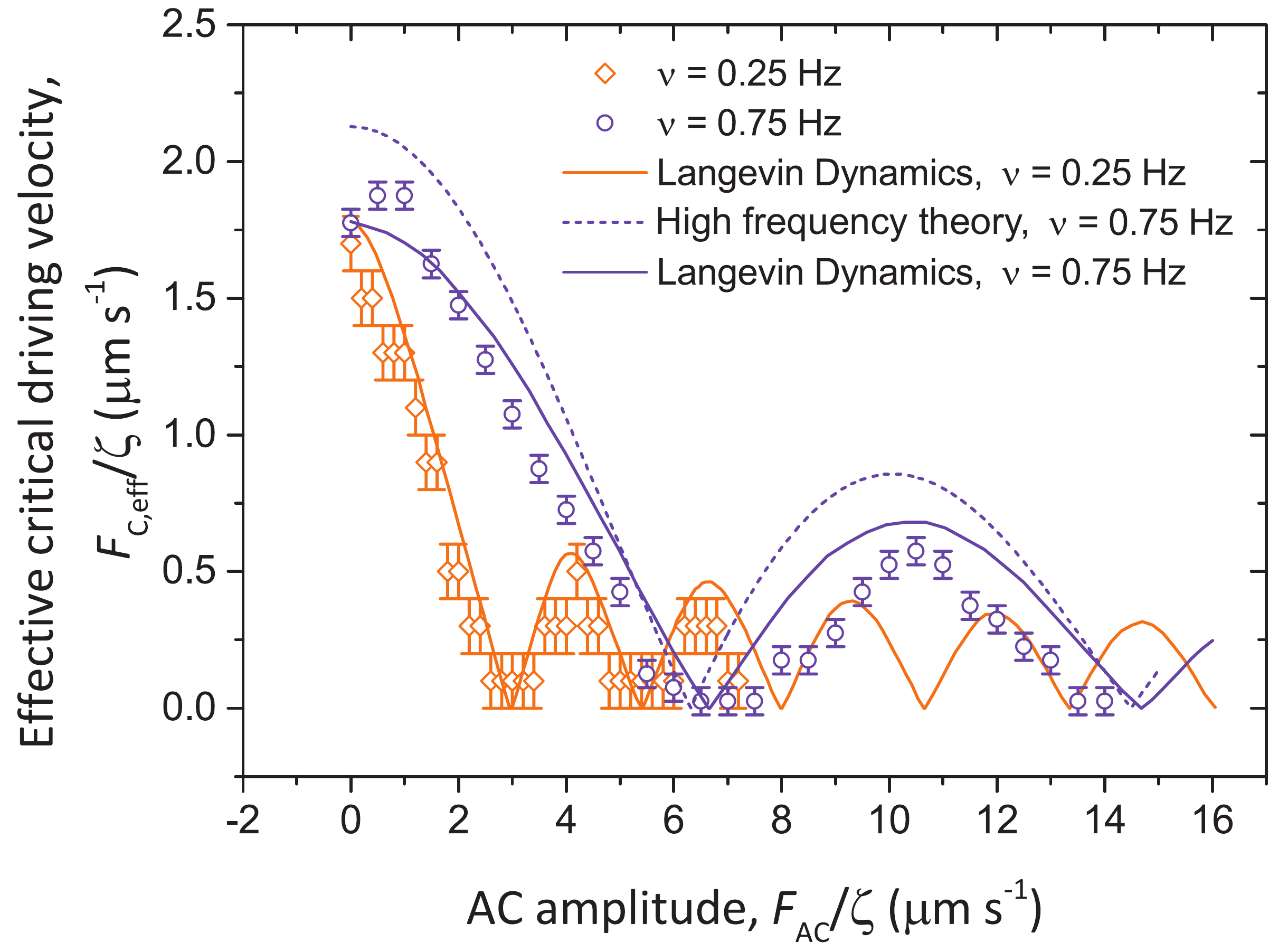}
	\caption{\small{Critical driving velocity, $F_{\textup{C,eff}}/\zeta$, as a function of modulation amplitude, $F_{\textup{AC}}/\zeta$, for $\nu=\frac{1}{4}$ Hz and $\nu=\frac{3}{4}$ Hz.  Symbols show experimental data, solid lines show LD simulation results, and the dashed line shows the high frequency approximation.}}
	\label{fcvsFAC}
\end{figure}

The mode locking footprint can also be seen in the effective diffusion coefficient: figure \ref{state_diag_075}(b) shows the same calculated lines as panel (a), overlayed on $D_{\textup{eff}}$, represented by a colour scale.  The locked states are clearly visible as the dark blue regions on the state diagram, where the effective diffusion coefficient drops dramatically as seen in figure \ref{centvarfig}(c).  The unlocked states range from blue, where the particle position is largely predictable, to the yellow and red regions between the mode locked steps.  An interesting observation may be made in the region between the zeroth and first modes, where the theory predicts no gap between the second critical line of the zeroth mode and the first critical line of the first mode.  The effective diffusion in this region is especially high, indicating that in the stochastic system the particle trajectory is highly unpredictable.  It is probable that in this region, the particle rapidly jumps between periods of being pinned to the landscape, and being in the first mode.

\subsubsection*{Critical driving velocity}

The state diagrams in figures \ref{state_diag_025} and \ref{state_diag_075} show that the critical driving velocity, $F_{\textup{C,eff}}$, oscillates as a function of modulation amplitude.  In figure \ref{fcvsFAC} the critical driving velocities are shown in isolation, and experimental results are compared to LD simulations and the high frequency theory.  Both sets of data, for $\nu=\frac{1}{4}$~Hz and $\nu=\frac{3}{4}$~Hz show a Bessel-function-like form, with the range of the oscillations determined by the frequency.  Each peak actually represents a different pinned mode \cite{Juniper2015}, and regions occur between the modes where the critical force is close to zero and thermal motion is sufficient to overcome the barriers.  The theoretical prediction for the high frequency regime (equation \ref{Fchighw}) predicts the shape of the experimental data reasonably well, but the LD simulations provide a somewhat better quantitative fit.

\subsection{Dependence on the frequency}
\label{freq}

\begin{figure}[t]
\centering
	\includegraphics[width=1\columnwidth]{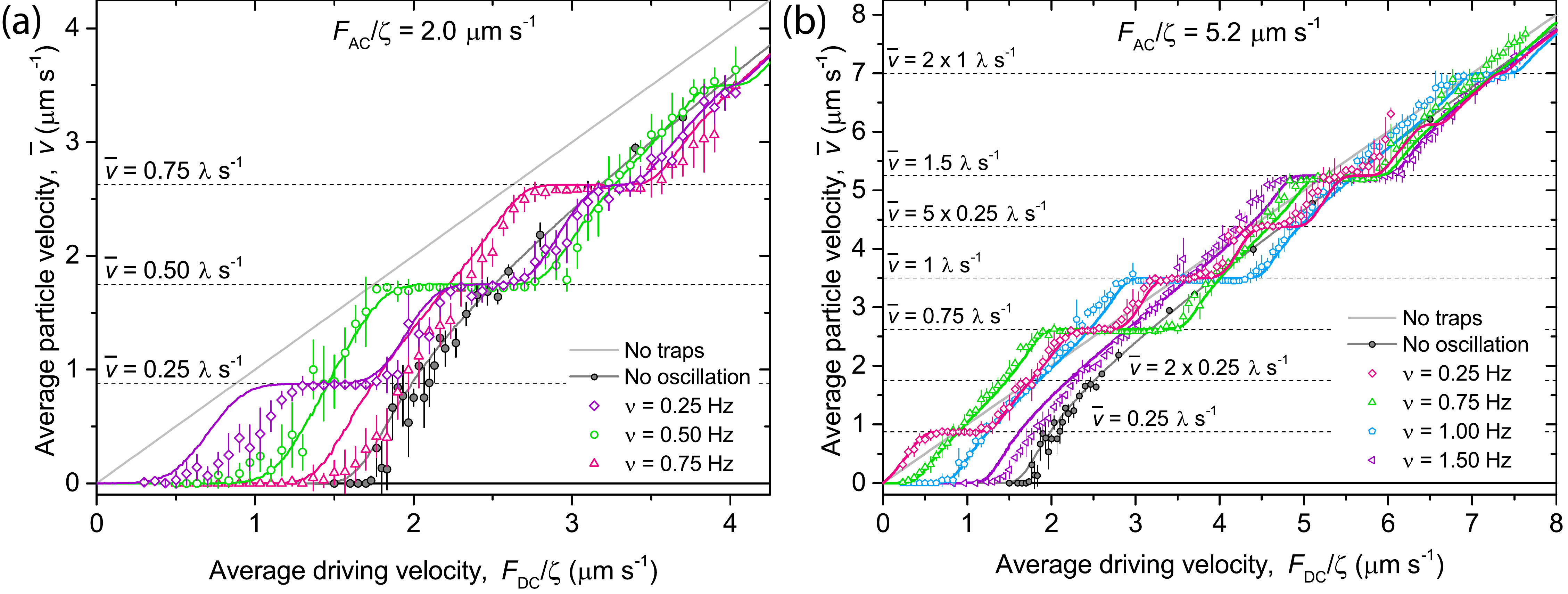}
	\caption{\small{Average particle velocity as a function of average driving velocity, at constant AC amplitude of (a) $F_{\textup{AC}}/\zeta=2.0~\upmu$m~s$^{-1}$ and (b) $F_{\textup{AC}}/\zeta=5.2~\upmu$m~s$^{-1}$.  Symbols show experimental data, solid lines show results of Langevin Dynamics simulations.  Dashed lines indicate the position of the first step for each frequency, and some higher $n$ steps for completeness.  Dark grey circles and line indicate case for no oscillation \cite{Juniper2016}, light grey line indicates case for no traps.}}
		\label{allfac2}
\end{figure}

The frequency dependence of the step velocities is expressed in the synchronisation condition, $\overline{v}=n\lambda\nu$.  Figure \ref{allfac2}(a) shows $\overline{v}$ as a function of $F_{\textup{DC}}/\zeta$ at three different frequencies, for an amplitude of $F_{\textup{AC}}/\zeta=2.0~\upmu$m~s$^{-1}$.  At a lower frequency more steps are seen over the same range of particle velocities as more harmonics are attainable.  Indeed at the lowest frequency, $\nu=\frac{1}{4}$~Hz, the mean particle velocity shows three steps corresponding to the first three integer multiples of $\nu\lambda$.  The second and third steps at this frequency therefore coincide with the first steps for the other two frequencies, of $\nu=\frac{1}{2}$~Hz and $\nu=\frac{3}{4}$~Hz.  As was seen in figure \ref{allw075}(a) and (b), there is generally a good agreement between the LD simulations and the experiments.

Figure \ref{allfac2}(b) shows the frequency dependence of the average particle velocity for an amplitude of $F_{\textup{AC}}/\zeta=5.2~\upmu$m~s$^{-1}$, for frequencies ranging from $\nu=\frac{1}{4}$ Hz to $\nu=\frac{3}{2}$ Hz.  It is worth noting that, as was seen in figure \ref{allw075}(b), not all possible steps appear; for example there is no visible step at $\overline{v}=\frac{1}{2} \lambda$ s$^{-1}$ for $\nu=\frac{1}{4}$ Hz.  Both figures \ref{allfac2}(a) and \ref{allfac2}(b) show that step width is frequency dependent, in addition to the amplitude dependence shown above.  Furthermore, the effective critical driving velocity clearly depends on the frequency: in both cases it is seen to increase with $\nu$, approaching the oscillation-free critical driving velocity of $F_{\textup{C}}/\zeta\approx1.8~\upmu$m~s$^{-1}$.

\begin{figure*}[t]
	\includegraphics[width=1.0\textwidth]{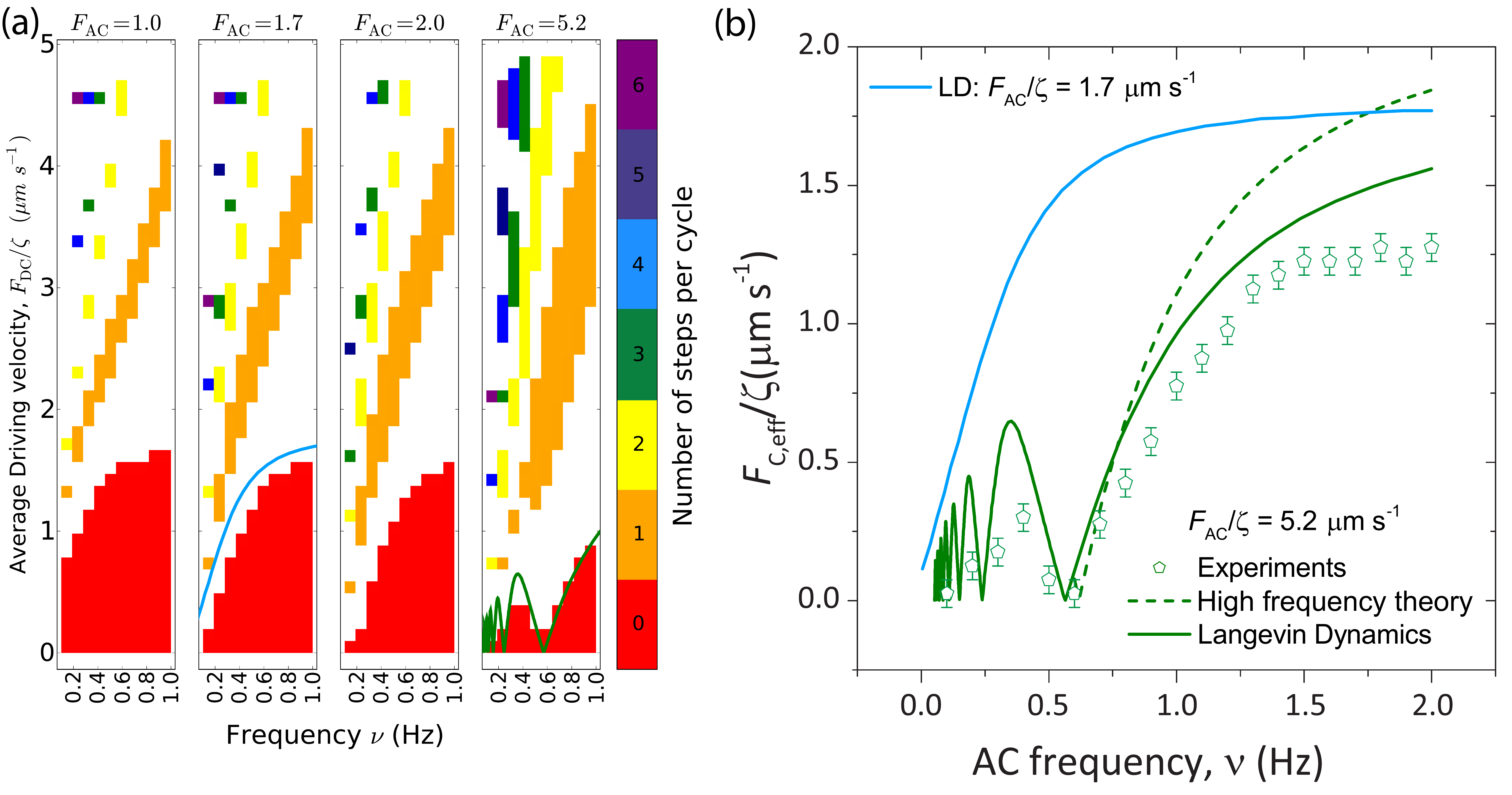}
	\centering
	\caption{\small{Frequency-dependence state diagrams and effective critical driving velocity.  (a) State diagrams show locked dynamic modes for fixed amplitudes of $F_{\textup{AC}}=1.0~\upmu$m~s$^{-1}$, $1.7~\upmu$m~s$^{-1}$, $2.0~\upmu$m~s$^{-1}$, and $5.2~\upmu$m~s$^{-1}$.  Coloured regions represent locked states as determined from DDFT, white regions are unlocked.  Locked states are numbered by an integer step number, $n$.  Solid lines show LD simulation results.  (b) Effective critical driving velocity, $F_{\textup{C,eff}}/\zeta$, as a function of frequency for $F_{\textup{AC}}/\zeta=5.2~\upmu$m~s$^{-1}$.  Symbols show experimental data, solid lines shows LD simulation results, dashed line shows high-frequency theory (equation \ref{Fchighw}).  Also shown in blue is the LD simulation result for $F_{\textup{AC}}=1.7~\upmu$m~s$^{-1}$.}}
	\label{state_diags_freq}
\end{figure*}

\subsubsection*{State diagrams and critical driving velocity}

DDFT computation is used to produce frequency-dependence state diagrams, which are not feasable experimentally as very low frequencies require extremely long run times.  Figure \ref{state_diags_freq}(a) shows four such state diagrams, at a range of amplitudes from $F_{\textup{AC}}< F_{\textup{C}}$ to $F_{\textup{AC}}> F_{\textup{C}}$.  Also included on two of the state diagrams are effective critical driving velocity lines determined from LD simulations.  As with the experiments, the run times required to produce further LD data, particularly at low frequencies, are prohibitively long.  From the data presented, however, the origin of the increasing effective critical driving velocity seen in figures \ref{allfac2}(a) and \ref{allfac2}(b) is clear, although the LD line on the  $F_{\textup{AC}}/\zeta=5.2~\upmu$m~s$^{-1}$ plot in particular highlights that the picture is more complex.  At $F_{\textup{AC}}>F_{\textup{C}}$ (where $F_{\textup{C}}$ is the DC critical driving velocity), a series of bumps appear at low frequency, which are too fine to be resolved by the DDFT data.  Below $F_{\textup{C}}$, however (i.e. at $F_{\textup{AC}}/\zeta=1.7~\upmu$m~s$^{-1}$), these bumps disappear, and the depinning transition is defined as a single monotonic increase.  This effect has been noted previously for the Frenkel Kontorova model, for a chain of interacting particles \cite{Hu2007,Tekic2008,TekicMali-book}.

Figure \ref{state_diags_freq}(b) shows the frequency dependent effective critical driving velocity for $F_{\textup{AC}}>F_{\textup{C}}$ in more detail.  Experimentally determined values for $F_{\textup{C}}$ are shown, along with the LD simulation result.  The experimental data shows a single smaller bump similar to that seen in the DDFT data, as it is also unable to resolve the numerous smaller bumps shown by the LD data.  The large bump at higher frequency, which plateaus at the DC-only critical driving velocity $F_{\textup{C}}$, represents the truly pinned state, where the particle does not move at all during a cycle of the oscillation, whereas the smaller bumps and lower frequency represent states where the \emph{net} particle motion is zero, as it moves but returns to the same potential well at the end of every cycle.  Also presented in figure \ref{state_diags_freq}(b) is the prediction of the high frequency approximation, which captures the form of the experimental data, but has a higher magnitude due to the absence of a noise term in equation \ref{Fchighw}, and the LD line for $F_{\textup{AC}}/\zeta=1.7~\upmu$m~s$^{-1}$, for comparison.

\section{Conclusions}
\label{mlconc}

Colloidal particles driven by the sum of constant and oscillating forces through a quasi-one-dimensional periodic optical potential energy landscape have been shown to exhibit rich non-linear dynamical behaviour.  Experiments showed that when an oscillating drive is applied, the average particle velocity has a staircase-like dependence on the average driving velocity, where the steps represent states of synchronisation between the particle motion and the substrate potential.  These results could be faithfuly reproduced using Langevin Dynamics simulations and Dynamic Density Functional Theory (DDFT), and in conditions of a high driving frequency the data mapped surprisingly well to an analytic approximation.  Probing the variance in the particle position and the trajectory diffusion using DDFT showed that the effective diffusion coefficient drops dramatically at the resonant mode-locked steps, explaining why the analytic theory (calculated at `zero temperature' with no fluctuations) is so successful.

The use of simulation and computation in addition to experimental results allowed a full exploration of the amplitude and frequency dependence of dynamic mode locking.  State diagrams showing both the amplitude and frequency dependence of the extent of the locked modes exposed the oscillating nature of the critical lines which define the mode locked steps.  These critical lines enclose regions which have been previously shown to represent different dynamic modes with the same net particle motion.  Finally, the effective critical driving velocity below which a particle is pinned to the potential landscape has been studied, and it has been shown to have an oscillating dependence on the modulation amplitude, with some conditions having no effective critical driving velocity.  The frequency dependence has been shown to be more complex, depending on whether the amplitude of the oscillation is above or below the critical driving velocity defined by the landscape.

By using a combination of experiments, computation, and analytic theory, it has been possible to explore the effect of a very wide range of conditions on dynamic mode locking, thereby giving a solid experimental and theoretical foundation to this dynamic synchronisation phenomenon.

\section*{Acknowledgements}
We thank Alice Thorneywork and Arran Curran for fruitful discussions. MPNJ, DGALA, and RPAD acknowledge EPSRC for financial support.  UZ acknowledges support from the German Academic Exchange Service (DAAD).

\section{Appendix}

\subsection{High frequency theory - further details}
\label{HighFreqAppendix}

In order to solve equation \ref{acdceom} for a certain range of driving frequencies, the work of Cotteverte et al. \cite{Cotteverte1994}, Chow et al. \cite{Chow1985}, and Reichhardt et al. \cite{Reichhardt2000} is followed.  Ideas developed in these previous works are used to draw an analytical approximation in this work. The first step in solving equation \ref{acdceom} is to neglect noise and split the particle trajectory, $x(t)$, into a part due to the terms independent of $x(t)$ and a deviation from it, caused by terms dependent on $x(t)$:
\begin{equation}\label{splitxt}
	x(t)=x_{\textup{0}}(t)+\delta (t).
\end{equation}
Accordingly, the equation for $x_{\textup{0}}(t)$ is taken to contain the DC and AC parts of the driving force,
\begin{equation}\label{xlin}
	\zeta\,\frac{\textup{d}x_{\textup{0}}(t)}{\textup{d}t}=F_{\textup{DC}}+F_{\textup{AC}}\cos\left(\omega t\right),
\end{equation}
which can be integrated to yield
\begin{equation}\label{x0(t)}
	\zeta \,x_{\textup{0}}(t)=F_{\textup{DC}}t+\frac{F_{\textup{AC}}}{\omega}\sin\left(\omega t\right).
\end{equation}
For the second part, $\delta(t)$, we obtain from equation \ref{acdceom}:
\begin{align}\label{ddeltatdt1}
	\zeta\,\frac{\textup{d}\delta (t)}{\textup{d}t} &= - F_{\textup{C}}\sin\left[\frac{2\uppi}{\lambda} \left(x_{\textup{0}}(t)+\delta(t)\right)\right] \nonumber\\
	&= - F_{\textup{C}}\sin\left[\frac{2\uppi}{\lambda}\left(\frac{F_{\textup{DC}}}{\zeta}t+\frac{F_{\textup{AC}}}{\zeta\omega}\sin\left(\omega t\right)+\delta(t)\right)\right],
\end{align}
where we have accounted for equations \ref{xlin} and \ref{x0(t)}.
Using the identity:
$
\sin\left[A\sin\left(\omega t\right)+B\right] \equiv \sum^{\infty}_{m=-\infty}J_m (A)\sin(B+m\omega t),
$
where $J_m$ is the $m$th order Bessel function of the first kind, $A=2\uppi F_{\textup{AC}}/(\zeta\omega \lambda)$ and $B=(2\uppi / \lambda) (F_{\textup{DC}}t/\zeta +\delta)$, equation \ref{ddeltatdt1} becomes:
\begin{align}\label{ddeltatdt}
	\zeta\,\frac{\textup{d}\delta (t)}{\textup{d}t} = - F_{\textup{C}}\sum^{\infty}_{m=-\infty}J_m \left(\frac{2\uppi}{\lambda}\frac{F_{\textup{AC}}}{\zeta\omega}\right)~\sin\left[\left(\frac{2\uppi}{\lambda}\frac{F_{\textup{DC}}}{\zeta} + m\omega \right)t + \frac{2\uppi}{\lambda}\delta (t)\right].
\end{align}
Equation \ref{ddeltatdt} is difficult to solve, so an approximation is made that only the leading term of the sum is retained, where 
\begin{align}\label{m-r}
m\omega = -r\omega \approx -\frac{2\uppi F_{\textup{DC}}}{\lambda\zeta} \quad (r=0,\pm 1,\pm 2,\ldots),
\end{align}
determines the mode locking. This approximation is valid at high enough frequencies. Indeed, the variation with time of the leading term with $m=-r$ given by equation \ref{m-r} is the slowest relative to $\omega t$, $2\omega t$, $\dots$ of the next to leading terms with $m=r\pm 1, r\pm 2, \ldots$. Provided that the dependence on $\delta(t)$ can be neglected in all next to leading terms, their averages over a period of the external modulation vanish. The scale of $\delta(t)$ can be estimated by noticing from equation \ref{ddeltatdt} that $\textup{d}\delta(t)/\textup{dt} \sim F_{\textup{C}}/\zeta$ (cf. \cite{Chow1985}) or, more accurately, $\textup{d}\delta(t)/\textup{dt} \sim J_{-r}(A) F_{\textup{C}}/\zeta$ and hence $\delta(t) \sim J_{-r}(A) F_{\textup{C}} t/\zeta$. By considering the term with the next to slowest variation, $m=-r\pm 1$, we require that $\omega t \gg \delta(t)$ to arrive at the condition for the validity of our high-frequency approximation:
\begin{equation}\label{condition}
	\omega \gg \frac{2\uppi}{\lambda}\frac{F_{\textup{C}}}{\zeta}J_{-r}\left(\frac{2\uppi}{\lambda}\frac{F_{\textup{AC}}}{\zeta\omega}\right) \quad \mbox{or} \quad \nu\gg\frac{F_{\textup{C}}}{\lambda\zeta} J_{-r}\left(\frac{F_{\textup{AC}}}{\lambda\nu\zeta}\right).
\end{equation}
Equation \ref{ddeltatdt} therefore becomes:
\begin{align}\label{ddeltatdt2}
	\zeta\frac{\textup{d}\delta(t)}{\textup{d}t} = -F_{\textup{C}}J_{-r}\left(\frac{F_{\textup{AC}}}{\lambda\nu\zeta}\right)\sin\left[ \frac{2\uppi}{\lambda}\left( \frac{\Delta F_{\textup{DC}}}{\zeta}t+\delta(t) \right) \right],
\end{align}
where $\Delta F_{\textup{DC}}=F_{\textup{DC}}-r\lambda\nu\zeta$ is a small change in the constant part of the driving force, $F_{\textup{DC}}$.  Introducing a variable $q(t)=(\Delta F_{\textup{DC}}/\zeta)t+\delta(t)$ further reduces equation \ref{ddeltatdt2} to the form of an Adler equation \cite{Adler1946, Goldstein2009}, equivalent to that found for the case of constant drive alone \cite{Juniper2016}:
\begin{equation}
	\zeta\frac{\textup{d}q(t)}{\textup{d}t}=\Delta F_{\textup{DC}}-F_{\textup{C}}J_{-r}\left( \frac{F_{\textup{AC}}}{\lambda\nu\zeta} \right) \sin\left[ \frac{2\uppi}{\lambda}q(t) \right].
\end{equation}
The expression for the average velocity may then be written, by noting that $\overline{v}=\left< \frac{\textup{d}x(t)}{\textup{d}t}\right>=F_{\textup{DC}}/\zeta+\left<\frac{\textup{d}\delta(t)}{\textup{d}t} \right>=r\lambda\nu+\left<\frac{\textup{d}q(t)}{\textup{d}t} \right>$, as the oscillating force term (the modulated part of the driving velocity) becomes zero after time averaging:

\begin{equation}\label{mlhighfreq_appendix}
	 \overline{v} = \left\{
\begin{aligned}&	r\lambda\nu, && \mbox{if }\quad |\Delta F_{\textup{DC}}| < \left|F_{\textup{C}} J_{-r}\left(\frac{F_{\textup{AC}}}{\lambda\nu\zeta}\right)\right|; \\
	& r\lambda\nu \pm \frac{1}{\zeta}\sqrt{\Delta F_{\textup{DC}}^2-F_{\textup{C}}^2 J_{-r}^2\left(\frac{F_{\textup{AC}}}{\lambda\nu\zeta}\right)}, && \mbox{if }\quad |\Delta F_{\textup{DC}}| > \left|F_{\textup{C}} J_{-r}\left(\frac{F_{\textup{AC}}}{\lambda\nu\zeta}\right)\right|, \quad \Delta F_{\textup{DC}} \gtrless 0, \end{aligned} \right.
\end{equation}
for $r=0,\pm 1,\pm 2,\ldots$.

\subsection{Dynamic Density Functional Theory - further details}
\label{DDFTAppendix}
{\bf Implementation details:}  The finite volume partial differential equation solver FiPy 3.1 \cite{FiPy:2009} is being used to perform the integration of the Smoluchowski equation. The grid of the computer system consists of 10,000 cells and has a total length of $50~\upmu$m ($\approx 14$ potential wells) with periodic boundary conditions.  The computations were terminated whenever the probability distribution was so widely spread that effects of periodicity could not be neglected.

{\bf Initial conditions:} The mean particle trajectory enters in general a short transient state before sychronising with the external AC driving force. This synchronised state is characterised by a periodic phase that modulates the linear drift of the mean trajectory. In order to suppress effects of the transient state we first estimate the mean position in the synchronised state of the respective system. Then we start the computation with a very narrow Gaussian function located in the determined position as the initial probability density $p(x, t=0)$.

{\bf Effective diffusion coefficient:}  In the mode locked states the limit of equation \ref{Deff} could not be reached within the time span of our computations due to intermediate plateaus as shown in figure \ref{centvarfig}(a).  In these cases the linear increase of the plateaus for 500 oscillations was calculated and used to determine $D_{\textup{eff}}$.

\bibliographystyle{apsrev4-1}
\bibliography{Bibliography.bib}

\begin{thebibliography}{72}%
\makeatletter
\providecommand \@ifxundefined [1]{%
 \@ifx{#1\undefined}
}%
\providecommand \@ifnum [1]{%
 \ifnum #1\expandafter \@firstoftwo
 \else \expandafter \@secondoftwo
 \fi
}%
\providecommand \@ifx [1]{%
 \ifx #1\expandafter \@firstoftwo
 \else \expandafter \@secondoftwo
 \fi
}%
\providecommand \natexlab [1]{#1}%
\providecommand \enquote  [1]{``#1''}%
\providecommand \bibnamefont  [1]{#1}%
\providecommand \bibfnamefont [1]{#1}%
\providecommand \citenamefont [1]{#1}%
\providecommand \href@noop [0]{\@secondoftwo}%
\providecommand \href [0]{\begingroup \@sanitize@url \@href}%
\providecommand \@href[1]{\@@startlink{#1}\@@href}%
\providecommand \@@href[1]{\endgroup#1\@@endlink}%
\providecommand \@sanitize@url [0]{\catcode `\\12\catcode `\$12\catcode
  `\&12\catcode `\#12\catcode `\^12\catcode `\_12\catcode `\%12\relax}%
\providecommand \@@startlink[1]{}%
\providecommand \@@endlink[0]{}%
\providecommand \url  [0]{\begingroup\@sanitize@url \@url }%
\providecommand \@url [1]{\endgroup\@href {#1}{\urlprefix }}%
\providecommand \urlprefix  [0]{URL }%
\providecommand \Eprint [0]{\href }%
\providecommand \doibase [0]{http://dx.doi.org/}%
\providecommand \selectlanguage [0]{\@gobble}%
\providecommand \bibinfo  [0]{\@secondoftwo}%
\providecommand \bibfield  [0]{\@secondoftwo}%
\providecommand \translation [1]{[#1]}%
\providecommand \BibitemOpen [0]{}%
\providecommand \bibitemStop [0]{}%
\providecommand \bibitemNoStop [0]{.\EOS\space}%
\providecommand \EOS [0]{\spacefactor3000\relax}%
\providecommand \BibitemShut  [1]{\csname bibitem#1\endcsname}%
\let\auto@bib@innerbib\@empty
\bibitem [{\citenamefont {Pikovsky}\ \emph {et~al.}(2001)\citenamefont
  {Pikovsky}, \citenamefont {Rosenblum},\ and\ \citenamefont
  {Kurths}}]{Pikovsky2001}%
  \BibitemOpen
  \bibfield  {author} {\bibinfo {author} {\bibfnamefont {A.}~\bibnamefont
  {Pikovsky}}, \bibinfo {author} {\bibfnamefont {M.}~\bibnamefont {Rosenblum}},
  \ and\ \bibinfo {author} {\bibfnamefont {J.}~\bibnamefont {Kurths}},\
  }\href@noop {} {\emph {\bibinfo {title} {Synchronization A universal concept
  in Nonlinear Sciences}}},\ Cambridge Nonlinear Science Series\ (\bibinfo
  {publisher} {Cambridge University Press},\ \bibinfo {address} {Cambridge},\
  \bibinfo {year} {2001})\BibitemShut {NoStop}%
\bibitem [{\citenamefont {Birch}(1756)}]{Birch1756}%
  \BibitemOpen
  \bibfield  {author} {\bibinfo {author} {\bibfnamefont {T.}~\bibnamefont
  {Birch}},\ }\href@noop {} {\emph {\bibinfo {title} {The History of the Royal
  Society of London}}}\ (\bibinfo {year} {1756})\BibitemShut {NoStop}%
\bibitem [{\citenamefont {Bennett}\ \emph {et~al.}(2002)\citenamefont
  {Bennett}, \citenamefont {Schatz}, \citenamefont {Rockwood},\ and\
  \citenamefont {Wiesenfeld}}]{Bennett2001}%
  \BibitemOpen
  \bibfield  {author} {\bibinfo {author} {\bibfnamefont {M.}~\bibnamefont
  {Bennett}}, \bibinfo {author} {\bibfnamefont {M.~F.}\ \bibnamefont {Schatz}},
  \bibinfo {author} {\bibfnamefont {H.}~\bibnamefont {Rockwood}}, \ and\
  \bibinfo {author} {\bibfnamefont {K.}~\bibnamefont {Wiesenfeld}},\
  }\href@noop {} {\bibfield  {journal} {\bibinfo  {journal} {Proceedings of the
  Royal Society A-Mathematical Physical and Engineering Sciences}\ }\textbf
  {\bibinfo {volume} {458}},\ \bibinfo {pages} {563} (\bibinfo {year}
  {2002})}\BibitemShut {NoStop}%
\bibitem [{\citenamefont {Agrawal}\ \emph {et~al.}(2013)\citenamefont
  {Agrawal}, \citenamefont {Woodhouse},\ and\ \citenamefont
  {Seshia}}]{Agrawal2013}%
  \BibitemOpen
  \bibfield  {author} {\bibinfo {author} {\bibfnamefont {D.~K.}\ \bibnamefont
  {Agrawal}}, \bibinfo {author} {\bibfnamefont {J.}~\bibnamefont {Woodhouse}},
  \ and\ \bibinfo {author} {\bibfnamefont {A.~A.}\ \bibnamefont {Seshia}},\
  }\href@noop {} {\bibfield  {journal} {\bibinfo  {journal} {Physical Review
  Letters}\ }\textbf {\bibinfo {volume} {111}},\ \bibinfo {pages} {084101}
  (\bibinfo {year} {2013})}\BibitemShut {NoStop}%
\bibitem [{\citenamefont {Neda}\ \emph
  {et~al.}(2000{\natexlab{a}})\citenamefont {Neda}, \citenamefont {Ravasz},
  \citenamefont {Brechet}, \citenamefont {Vicsek},\ and\ \citenamefont
  {Barabasi}}]{Neda2000}%
  \BibitemOpen
  \bibfield  {author} {\bibinfo {author} {\bibfnamefont {Z.}~\bibnamefont
  {Neda}}, \bibinfo {author} {\bibfnamefont {E.}~\bibnamefont {Ravasz}},
  \bibinfo {author} {\bibfnamefont {Y.}~\bibnamefont {Brechet}}, \bibinfo
  {author} {\bibfnamefont {T.}~\bibnamefont {Vicsek}}, \ and\ \bibinfo {author}
  {\bibfnamefont {A.~L.}\ \bibnamefont {Barabasi}},\ }\href@noop {} {\bibfield
  {journal} {\bibinfo  {journal} {Nature}\ }\textbf {\bibinfo {volume} {403}},\
  \bibinfo {pages} {849} (\bibinfo {year} {2000}{\natexlab{a}})}\BibitemShut
  {NoStop}%
\bibitem [{\citenamefont {Neda}\ \emph
  {et~al.}(2000{\natexlab{b}})\citenamefont {Neda}, \citenamefont {Ravasz},
  \citenamefont {Vicsek}, \citenamefont {Brechet},\ and\ \citenamefont
  {Barabasi}}]{Neda2000PRE}%
  \BibitemOpen
  \bibfield  {author} {\bibinfo {author} {\bibfnamefont {Z.}~\bibnamefont
  {Neda}}, \bibinfo {author} {\bibfnamefont {E.}~\bibnamefont {Ravasz}},
  \bibinfo {author} {\bibfnamefont {T.}~\bibnamefont {Vicsek}}, \bibinfo
  {author} {\bibfnamefont {Y.}~\bibnamefont {Brechet}}, \ and\ \bibinfo
  {author} {\bibfnamefont {A.~L.}\ \bibnamefont {Barabasi}},\ }\href@noop {}
  {\bibfield  {journal} {\bibinfo  {journal} {Physical Review E}\ }\textbf
  {\bibinfo {volume} {61}},\ \bibinfo {pages} {6987} (\bibinfo {year}
  {2000}{\natexlab{b}})}\BibitemShut {NoStop}%
\bibitem [{\citenamefont {Nicolis}\ \emph {et~al.}(2013)\citenamefont
  {Nicolis}, \citenamefont {Fernández}, \citenamefont {Pérez-Penichet},
  \citenamefont {Noda}, \citenamefont {Tejera}, \citenamefont {Ramos},
  \citenamefont {Sumpter},\ and\ \citenamefont {Altshuler}}]{Nicolis2013}%
  \BibitemOpen
  \bibfield  {author} {\bibinfo {author} {\bibfnamefont {S.~C.}\ \bibnamefont
  {Nicolis}}, \bibinfo {author} {\bibfnamefont {J.}~\bibnamefont {Fernández}},
  \bibinfo {author} {\bibfnamefont {C.}~\bibnamefont {Pérez-Penichet}},
  \bibinfo {author} {\bibfnamefont {C.}~\bibnamefont {Noda}}, \bibinfo {author}
  {\bibfnamefont {F.}~\bibnamefont {Tejera}}, \bibinfo {author} {\bibfnamefont
  {O.}~\bibnamefont {Ramos}}, \bibinfo {author} {\bibfnamefont {D.~J.~T.}\
  \bibnamefont {Sumpter}}, \ and\ \bibinfo {author} {\bibfnamefont
  {E.}~\bibnamefont {Altshuler}},\ }\href@noop {} {\bibfield  {journal}
  {\bibinfo  {journal} {Physical Review Letters}\ }\textbf {\bibinfo {volume}
  {110}},\ \bibinfo {pages} {268104} (\bibinfo {year} {2013})}\BibitemShut
  {NoStop}%
\bibitem [{\citenamefont {Feng}\ \emph {et~al.}(2008)\citenamefont {Feng},
  \citenamefont {White}, \citenamefont {Hajimiri},\ and\ \citenamefont
  {Roukes}}]{Feng2008}%
  \BibitemOpen
  \bibfield  {author} {\bibinfo {author} {\bibfnamefont {X.~L.}\ \bibnamefont
  {Feng}}, \bibinfo {author} {\bibfnamefont {C.~J.}\ \bibnamefont {White}},
  \bibinfo {author} {\bibfnamefont {A.}~\bibnamefont {Hajimiri}}, \ and\
  \bibinfo {author} {\bibfnamefont {M.~L.}\ \bibnamefont {Roukes}},\
  }\href@noop {} {\bibfield  {journal} {\bibinfo  {journal} {Nature
  Nanotechnology}\ }\textbf {\bibinfo {volume} {3}},\ \bibinfo {pages} {342}
  (\bibinfo {year} {2008})}\BibitemShut {NoStop}%
\bibitem [{\citenamefont {Antonio}\ \emph {et~al.}(2012)\citenamefont
  {Antonio}, \citenamefont {Zanette},\ and\ \citenamefont
  {Lopez}}]{Antonio2012}%
  \BibitemOpen
  \bibfield  {author} {\bibinfo {author} {\bibfnamefont {D.}~\bibnamefont
  {Antonio}}, \bibinfo {author} {\bibfnamefont {D.~H.}\ \bibnamefont
  {Zanette}}, \ and\ \bibinfo {author} {\bibfnamefont {D.}~\bibnamefont
  {Lopez}},\ }\href@noop {} {\bibfield  {journal} {\bibinfo  {journal} {Nature
  Communications}\ }\textbf {\bibinfo {volume} {3}},\ \bibinfo {pages} {806}
  (\bibinfo {year} {2012})}\BibitemShut {NoStop}%
\bibitem [{\citenamefont {Shim}\ \emph {et~al.}(2007)\citenamefont {Shim},
  \citenamefont {Imboden},\ and\ \citenamefont {Mohanty}}]{Shim2007}%
  \BibitemOpen
  \bibfield  {author} {\bibinfo {author} {\bibfnamefont {S.~B.}\ \bibnamefont
  {Shim}}, \bibinfo {author} {\bibfnamefont {M.}~\bibnamefont {Imboden}}, \
  and\ \bibinfo {author} {\bibfnamefont {P.}~\bibnamefont {Mohanty}},\
  }\href@noop {} {\bibfield  {journal} {\bibinfo  {journal} {Science}\ }\textbf
  {\bibinfo {volume} {316}},\ \bibinfo {pages} {95} (\bibinfo {year}
  {2007})}\BibitemShut {NoStop}%
\bibitem [{\citenamefont {Zalalutdinov}\ \emph {et~al.}(2003)\citenamefont
  {Zalalutdinov}, \citenamefont {Aubin}, \citenamefont {Pandey}, \citenamefont
  {Zehnder}, \citenamefont {Rand}, \citenamefont {Craighead}, \citenamefont
  {Parpia},\ and\ \citenamefont {Houston}}]{Zalalutdinov2003}%
  \BibitemOpen
  \bibfield  {author} {\bibinfo {author} {\bibfnamefont {M.}~\bibnamefont
  {Zalalutdinov}}, \bibinfo {author} {\bibfnamefont {K.~L.}\ \bibnamefont
  {Aubin}}, \bibinfo {author} {\bibfnamefont {M.}~\bibnamefont {Pandey}},
  \bibinfo {author} {\bibfnamefont {A.~T.}\ \bibnamefont {Zehnder}}, \bibinfo
  {author} {\bibfnamefont {R.~H.}\ \bibnamefont {Rand}}, \bibinfo {author}
  {\bibfnamefont {H.~G.}\ \bibnamefont {Craighead}}, \bibinfo {author}
  {\bibfnamefont {J.~M.}\ \bibnamefont {Parpia}}, \ and\ \bibinfo {author}
  {\bibfnamefont {B.~H.}\ \bibnamefont {Houston}},\ }\href@noop {} {\bibfield
  {journal} {\bibinfo  {journal} {Applied Physics Letters}\ }\textbf {\bibinfo
  {volume} {83}},\ \bibinfo {pages} {3281} (\bibinfo {year}
  {2003})}\BibitemShut {NoStop}%
\bibitem [{\citenamefont {Zhang}\ \emph {et~al.}(2012)\citenamefont {Zhang},
  \citenamefont {Wiederhecker}, \citenamefont {Manipatruni}, \citenamefont
  {Barnard}, \citenamefont {McEuen},\ and\ \citenamefont {Lipson}}]{Zhang2012}%
  \BibitemOpen
  \bibfield  {author} {\bibinfo {author} {\bibfnamefont {M.~A.}\ \bibnamefont
  {Zhang}}, \bibinfo {author} {\bibfnamefont {G.~S.}\ \bibnamefont
  {Wiederhecker}}, \bibinfo {author} {\bibfnamefont {S.}~\bibnamefont
  {Manipatruni}}, \bibinfo {author} {\bibfnamefont {A.}~\bibnamefont
  {Barnard}}, \bibinfo {author} {\bibfnamefont {P.}~\bibnamefont {McEuen}}, \
  and\ \bibinfo {author} {\bibfnamefont {M.}~\bibnamefont {Lipson}},\
  }\href@noop {} {\bibfield  {journal} {\bibinfo  {journal} {Physical Review
  Letters}\ }\textbf {\bibinfo {volume} {109}},\ \bibinfo {pages} {233906}
  (\bibinfo {year} {2012})}\BibitemShut {NoStop}%
\bibitem [{\citenamefont {Besseling}\ \emph {et~al.}(2005)\citenamefont
  {Besseling}, \citenamefont {Kes}, \citenamefont {Drose},\ and\ \citenamefont
  {Vinokur}}]{Besseling2005}%
  \BibitemOpen
  \bibfield  {author} {\bibinfo {author} {\bibfnamefont {R.}~\bibnamefont
  {Besseling}}, \bibinfo {author} {\bibfnamefont {P.~H.}\ \bibnamefont {Kes}},
  \bibinfo {author} {\bibfnamefont {T.}~\bibnamefont {Drose}}, \ and\ \bibinfo
  {author} {\bibfnamefont {V.~M.}\ \bibnamefont {Vinokur}},\ }\href@noop {}
  {\bibfield  {journal} {\bibinfo  {journal} {New Journal of Physics}\ }\textbf
  {\bibinfo {volume} {7}},\ \bibinfo {pages} {71} (\bibinfo {year}
  {2005})}\BibitemShut {NoStop}%
\bibitem [{\citenamefont {Olson}\ \emph {et~al.}(1998)\citenamefont {Olson},
  \citenamefont {Reichhardt},\ and\ \citenamefont {Nori}}]{Olsen1998}%
  \BibitemOpen
  \bibfield  {author} {\bibinfo {author} {\bibfnamefont {C.~J.}\ \bibnamefont
  {Olson}}, \bibinfo {author} {\bibfnamefont {C.}~\bibnamefont {Reichhardt}}, \
  and\ \bibinfo {author} {\bibfnamefont {F.}~\bibnamefont {Nori}},\ }\href@noop
  {} {\bibfield  {journal} {\bibinfo  {journal} {Physical Review Letters}\
  }\textbf {\bibinfo {volume} {81}},\ \bibinfo {pages} {3757} (\bibinfo {year}
  {1998})}\BibitemShut {NoStop}%
\bibitem [{\citenamefont {Kokubo}\ \emph {et~al.}(2004)\citenamefont {Kokubo},
  \citenamefont {Besseling},\ and\ \citenamefont {Kes}}]{Kokubo2004}%
  \BibitemOpen
  \bibfield  {author} {\bibinfo {author} {\bibfnamefont {N.}~\bibnamefont
  {Kokubo}}, \bibinfo {author} {\bibfnamefont {R.}~\bibnamefont {Besseling}}, \
  and\ \bibinfo {author} {\bibfnamefont {P.~H.}\ \bibnamefont {Kes}},\
  }\href@noop {} {\bibfield  {journal} {\bibinfo  {journal} {Physical Review
  B}\ }\textbf {\bibinfo {volume} {69}},\ \bibinfo {pages} {064504} (\bibinfo
  {year} {2004})}\BibitemShut {NoStop}%
\bibitem [{\citenamefont {Kokubo}\ \emph {et~al.}(2002)\citenamefont {Kokubo},
  \citenamefont {Besseling}, \citenamefont {Vinokur},\ and\ \citenamefont
  {Kes}}]{Kokubo2002}%
  \BibitemOpen
  \bibfield  {author} {\bibinfo {author} {\bibfnamefont {N.}~\bibnamefont
  {Kokubo}}, \bibinfo {author} {\bibfnamefont {R.}~\bibnamefont {Besseling}},
  \bibinfo {author} {\bibfnamefont {V.~M.}\ \bibnamefont {Vinokur}}, \ and\
  \bibinfo {author} {\bibfnamefont {P.~H.}\ \bibnamefont {Kes}},\ }\href@noop
  {} {\bibfield  {journal} {\bibinfo  {journal} {Physical Review Letters}\
  }\textbf {\bibinfo {volume} {88}},\ \bibinfo {pages} {247004} (\bibinfo
  {year} {2002})}\BibitemShut {NoStop}%
\bibitem [{\citenamefont {Kolton}\ \emph {et~al.}(2001)\citenamefont {Kolton},
  \citenamefont {Dominguez},\ and\ \citenamefont
  {Gronbech-Jensen}}]{Kolton2001}%
  \BibitemOpen
  \bibfield  {author} {\bibinfo {author} {\bibfnamefont {A.~B.}\ \bibnamefont
  {Kolton}}, \bibinfo {author} {\bibfnamefont {D.}~\bibnamefont {Dominguez}}, \
  and\ \bibinfo {author} {\bibfnamefont {N.}~\bibnamefont {Gronbech-Jensen}},\
  }\href@noop {} {\bibfield  {journal} {\bibinfo  {journal} {Physical Review
  Letters}\ }\textbf {\bibinfo {volume} {86}},\ \bibinfo {pages} {4112}
  (\bibinfo {year} {2001})}\BibitemShut {NoStop}%
\bibitem [{\citenamefont {Grigorieva}(1994)}]{Grigorieva1994}%
  \BibitemOpen
  \bibfield  {author} {\bibinfo {author} {\bibfnamefont {I.~V.}\ \bibnamefont
  {Grigorieva}},\ }\href@noop {} {\bibfield  {journal} {\bibinfo  {journal}
  {Superconductor Science \& Technology}\ }\textbf {\bibinfo {volume} {7}},\
  \bibinfo {pages} {161} (\bibinfo {year} {1994})}\BibitemShut {NoStop}%
\bibitem [{\citenamefont {Aubry}(1978)}]{Aubry1978}%
  \BibitemOpen
  \bibfield  {author} {\bibinfo {author} {\bibfnamefont {S.}~\bibnamefont
  {Aubry}},\ }\enquote {\bibinfo {title} {The new concept of transitions by
  breaking of analyticity in a crystallographic model},}\ in\ \href@noop {}
  {\emph {\bibinfo {booktitle} {Solitons and Condensed Matter Physics}}},\
  \bibinfo {series and number} {Springer Series in Solid-State Sciences},\
  \bibinfo {editor} {edited by\ \bibinfo {editor} {\bibfnamefont
  {T.}~\bibnamefont {Bishop}, \bibfnamefont {A.~R.;~Schneider}}}\ (\bibinfo
  {publisher} {Springer-Verlag},\ \bibinfo {address} {Berlin},\ \bibinfo {year}
  {1978})\ pp.\ \bibinfo {pages} {264--277}\BibitemShut {NoStop}%
\bibitem [{\citenamefont {Talkner}\ \emph {et~al.}(1999)\citenamefont
  {Talkner}, \citenamefont {Hershkovitz}, \citenamefont {Pollak},\ and\
  \citenamefont {Hanggi}}]{Talkner1999}%
  \BibitemOpen
  \bibfield  {author} {\bibinfo {author} {\bibfnamefont {P.}~\bibnamefont
  {Talkner}}, \bibinfo {author} {\bibfnamefont {E.}~\bibnamefont
  {Hershkovitz}}, \bibinfo {author} {\bibfnamefont {E.}~\bibnamefont {Pollak}},
  \ and\ \bibinfo {author} {\bibfnamefont {P.}~\bibnamefont {Hanggi}},\
  }\href@noop {} {\bibfield  {journal} {\bibinfo  {journal} {Surface Science}\
  }\textbf {\bibinfo {volume} {437}},\ \bibinfo {pages} {198} (\bibinfo {year}
  {1999})}\BibitemShut {NoStop}%
\bibitem [{\citenamefont {Kvale}\ and\ \citenamefont
  {Hebboul}(1991)}]{Kvale1991}%
  \BibitemOpen
  \bibfield  {author} {\bibinfo {author} {\bibfnamefont {M.}~\bibnamefont
  {Kvale}}\ and\ \bibinfo {author} {\bibfnamefont {S.~E.}\ \bibnamefont
  {Hebboul}},\ }\href@noop {} {\bibfield  {journal} {\bibinfo  {journal}
  {Physical Review B}\ }\textbf {\bibinfo {volume} {43}},\ \bibinfo {pages}
  {3720} (\bibinfo {year} {1991})}\BibitemShut {NoStop}%
\bibitem [{\citenamefont {Burkov}(1991)}]{Burkov1991}%
  \BibitemOpen
  \bibfield  {author} {\bibinfo {author} {\bibfnamefont {S.~E.}\ \bibnamefont
  {Burkov}},\ }\href@noop {} {\bibfield  {journal} {\bibinfo  {journal}
  {Physical Review B}\ }\textbf {\bibinfo {volume} {44}},\ \bibinfo {pages}
  {2850} (\bibinfo {year} {1991})}\BibitemShut {NoStop}%
\bibitem [{\citenamefont {Grimes}\ and\ \citenamefont
  {Shapiro}(1968)}]{Grimes1968}%
  \BibitemOpen
  \bibfield  {author} {\bibinfo {author} {\bibfnamefont {C.~C.}\ \bibnamefont
  {Grimes}}\ and\ \bibinfo {author} {\bibfnamefont {S.}~\bibnamefont
  {Shapiro}},\ }\href@noop {} {\bibfield  {journal} {\bibinfo  {journal}
  {Physical Review}\ }\textbf {\bibinfo {volume} {169}},\ \bibinfo {pages}
  {397} (\bibinfo {year} {1968})}\BibitemShut {NoStop}%
\bibitem [{\citenamefont {Josephson}(1962)}]{Josephson1962}%
  \BibitemOpen
  \bibfield  {author} {\bibinfo {author} {\bibfnamefont {B.~D.}\ \bibnamefont
  {Josephson}},\ }\href@noop {} {\bibfield  {journal} {\bibinfo  {journal}
  {Physics Letters}\ }\textbf {\bibinfo {volume} {1}},\ \bibinfo {pages} {251}
  (\bibinfo {year} {1962})}\BibitemShut {NoStop}%
\bibitem [{\citenamefont {Gr\"{u}ner}(1988)}]{Gruner1988}%
  \BibitemOpen
  \bibfield  {author} {\bibinfo {author} {\bibfnamefont {G.}~\bibnamefont
  {Gr\"{u}ner}},\ }\href@noop {} {\bibfield  {journal} {\bibinfo  {journal}
  {Reviews of Modern Physics}\ }\textbf {\bibinfo {volume} {60}},\ \bibinfo
  {pages} {1129} (\bibinfo {year} {1988})}\BibitemShut {NoStop}%
\bibitem [{\citenamefont {Carpinelli}\ \emph {et~al.}(1996)\citenamefont
  {Carpinelli}, \citenamefont {Weitering}, \citenamefont {Plummer},\ and\
  \citenamefont {Stumpf}}]{Carpinelli1996}%
  \BibitemOpen
  \bibfield  {author} {\bibinfo {author} {\bibfnamefont {J.~M.}\ \bibnamefont
  {Carpinelli}}, \bibinfo {author} {\bibfnamefont {H.~H.}\ \bibnamefont
  {Weitering}}, \bibinfo {author} {\bibfnamefont {E.~W.}\ \bibnamefont
  {Plummer}}, \ and\ \bibinfo {author} {\bibfnamefont {R.}~\bibnamefont
  {Stumpf}},\ }\href@noop {} {\bibfield  {journal} {\bibinfo  {journal}
  {Nature}\ }\textbf {\bibinfo {volume} {381}},\ \bibinfo {pages} {398}
  (\bibinfo {year} {1996})}\BibitemShut {NoStop}%
\bibitem [{\citenamefont {Eichberger}\ \emph {et~al.}(2010)\citenamefont
  {Eichberger}, \citenamefont {Schafer}, \citenamefont {Krumova}, \citenamefont
  {Beyer}, \citenamefont {Demsar}, \citenamefont {Berger}, \citenamefont
  {Moriena}, \citenamefont {Sciaini},\ and\ \citenamefont
  {Miller}}]{Eichberger2010}%
  \BibitemOpen
  \bibfield  {author} {\bibinfo {author} {\bibfnamefont {M.}~\bibnamefont
  {Eichberger}}, \bibinfo {author} {\bibfnamefont {H.}~\bibnamefont {Schafer}},
  \bibinfo {author} {\bibfnamefont {M.}~\bibnamefont {Krumova}}, \bibinfo
  {author} {\bibfnamefont {M.}~\bibnamefont {Beyer}}, \bibinfo {author}
  {\bibfnamefont {J.}~\bibnamefont {Demsar}}, \bibinfo {author} {\bibfnamefont
  {H.}~\bibnamefont {Berger}}, \bibinfo {author} {\bibfnamefont
  {G.}~\bibnamefont {Moriena}}, \bibinfo {author} {\bibfnamefont
  {G.}~\bibnamefont {Sciaini}}, \ and\ \bibinfo {author} {\bibfnamefont
  {R.~J.~D.}\ \bibnamefont {Miller}},\ }\href@noop {} {\bibfield  {journal}
  {\bibinfo  {journal} {Nature}\ }\textbf {\bibinfo {volume} {468}},\ \bibinfo
  {pages} {799} (\bibinfo {year} {2010})}\BibitemShut {NoStop}%
\bibitem [{\citenamefont {Sancho}\ and\ \citenamefont
  {Lacasta}(2010)}]{Sancho2010}%
  \BibitemOpen
  \bibfield  {author} {\bibinfo {author} {\bibfnamefont {J.~M.}\ \bibnamefont
  {Sancho}}\ and\ \bibinfo {author} {\bibfnamefont {A.~M.}\ \bibnamefont
  {Lacasta}},\ }\href@noop {} {\bibfield  {journal} {\bibinfo  {journal} {The
  European Physical Journal Special Topics}\ }\textbf {\bibinfo {volume}
  {187}},\ \bibinfo {pages} {49} (\bibinfo {year} {2010})}\BibitemShut
  {NoStop}%
\bibitem [{\citenamefont {Dobnikar}\ \emph {et~al.}(2013)\citenamefont
  {Dobnikar}, \citenamefont {Snezhko},\ and\ \citenamefont
  {Yethiraj}}]{Dobnikar2013}%
  \BibitemOpen
  \bibfield  {author} {\bibinfo {author} {\bibfnamefont {J.}~\bibnamefont
  {Dobnikar}}, \bibinfo {author} {\bibfnamefont {A.}~\bibnamefont {Snezhko}}, \
  and\ \bibinfo {author} {\bibfnamefont {A.}~\bibnamefont {Yethiraj}},\
  }\href@noop {} {\bibfield  {journal} {\bibinfo  {journal} {Soft Matter}\
  }\textbf {\bibinfo {volume} {9}},\ \bibinfo {pages} {3693} (\bibinfo {year}
  {2013})}\BibitemShut {NoStop}%
\bibitem [{\citenamefont {Simon}\ and\ \citenamefont
  {Libchaber}(1992)}]{Simon1992}%
  \BibitemOpen
  \bibfield  {author} {\bibinfo {author} {\bibfnamefont {A.}~\bibnamefont
  {Simon}}\ and\ \bibinfo {author} {\bibfnamefont {A.}~\bibnamefont
  {Libchaber}},\ }\href@noop {} {\bibfield  {journal} {\bibinfo  {journal}
  {Physical Review Letters}\ }\textbf {\bibinfo {volume} {68}},\ \bibinfo
  {pages} {3375} (\bibinfo {year} {1992})}\BibitemShut {NoStop}%
\bibitem [{\citenamefont {Curran}\ \emph {et~al.}(2012)\citenamefont {Curran},
  \citenamefont {Lee}, \citenamefont {Leonardo},\ and\ \citenamefont
  {Padgett}}]{Curran2012}%
  \BibitemOpen
  \bibfield  {author} {\bibinfo {author} {\bibfnamefont {A.}~\bibnamefont
  {Curran}}, \bibinfo {author} {\bibfnamefont {M.~P.}\ \bibnamefont {Lee}},
  \bibinfo {author} {\bibfnamefont {R.~D.}\ \bibnamefont {Leonardo}}, \ and\
  \bibinfo {author} {\bibfnamefont {J.~M. C. M.~J.}\ \bibnamefont {Padgett}},\
  }\href@noop {} {\bibfield  {journal} {\bibinfo  {journal} {Physical Review
  Letters}\ }\textbf {\bibinfo {volume} {108}},\ \bibinfo {pages} {240601}
  (\bibinfo {year} {2012})}\BibitemShut {NoStop}%
\bibitem [{\citenamefont {Schmitt}\ \emph {et~al.}(2006)\citenamefont
  {Schmitt}, \citenamefont {Dybiec}, \citenamefont {H{\"a}nggi},\ and\
  \citenamefont {Bechinger}}]{Schmitt2006}%
  \BibitemOpen
  \bibfield  {author} {\bibinfo {author} {\bibfnamefont {C.}~\bibnamefont
  {Schmitt}}, \bibinfo {author} {\bibfnamefont {B.}~\bibnamefont {Dybiec}},
  \bibinfo {author} {\bibfnamefont {P.}~\bibnamefont {H{\"a}nggi}}, \ and\
  \bibinfo {author} {\bibfnamefont {C.}~\bibnamefont {Bechinger}},\ }\href@noop
  {} {\bibfield  {journal} {\bibinfo  {journal} {Europhysics Letters}\ }\textbf
  {\bibinfo {volume} {74}},\ \bibinfo {pages} {937} (\bibinfo {year}
  {2006})}\BibitemShut {NoStop}%
\bibitem [{\citenamefont {Babic}\ \emph {et~al.}(2004)\citenamefont {Babic},
  \citenamefont {Schmitt}, \citenamefont {Poberaj},\ and\ \citenamefont
  {Bechinger}}]{Babic2004}%
  \BibitemOpen
  \bibfield  {author} {\bibinfo {author} {\bibfnamefont {D.}~\bibnamefont
  {Babic}}, \bibinfo {author} {\bibfnamefont {C.}~\bibnamefont {Schmitt}},
  \bibinfo {author} {\bibfnamefont {I.}~\bibnamefont {Poberaj}}, \ and\
  \bibinfo {author} {\bibfnamefont {C.}~\bibnamefont {Bechinger}},\ }\href@noop
  {} {\bibfield  {journal} {\bibinfo  {journal} {Europhysics Letters}\ }\textbf
  {\bibinfo {volume} {67}},\ \bibinfo {pages} {158} (\bibinfo {year}
  {2004})}\BibitemShut {NoStop}%
\bibitem [{\citenamefont {Hennig}\ \emph {et~al.}(2010)\citenamefont {Hennig},
  \citenamefont {Burbanks},\ and\ \citenamefont {Osbaldestin}}]{Hennig2010}%
  \BibitemOpen
  \bibfield  {author} {\bibinfo {author} {\bibfnamefont {D.}~\bibnamefont
  {Hennig}}, \bibinfo {author} {\bibfnamefont {A.~D.}\ \bibnamefont
  {Burbanks}}, \ and\ \bibinfo {author} {\bibfnamefont {A.~H.}\ \bibnamefont
  {Osbaldestin}},\ }\href@noop {} {\bibfield  {journal} {\bibinfo  {journal}
  {Chemical Physics}\ }\textbf {\bibinfo {volume} {375}},\ \bibinfo {pages}
  {492} (\bibinfo {year} {2010})}\BibitemShut {NoStop}%
\bibitem [{\citenamefont {Hasnain}\ \emph {et~al.}(2013)\citenamefont
  {Hasnain}, \citenamefont {Jungblut},\ and\ \citenamefont
  {Dellago}}]{Hasnain2013}%
  \BibitemOpen
  \bibfield  {author} {\bibinfo {author} {\bibfnamefont {J.}~\bibnamefont
  {Hasnain}}, \bibinfo {author} {\bibfnamefont {S.}~\bibnamefont {Jungblut}}, \
  and\ \bibinfo {author} {\bibfnamefont {C.}~\bibnamefont {Dellago}},\
  }\href@noop {} {\bibfield  {journal} {\bibinfo  {journal} {Soft Matter}\
  }\textbf {\bibinfo {volume} {9}},\ \bibinfo {pages} {5867} (\bibinfo {year}
  {2013})}\BibitemShut {NoStop}%
\bibitem [{\citenamefont {McDermott}\ \emph {et~al.}(2013)\citenamefont
  {McDermott}, \citenamefont {Amelang}, \citenamefont {Lopatina}, \citenamefont
  {Reichhardt},\ and\ \citenamefont {Reichhardt}}]{Mcdermott2013}%
  \BibitemOpen
  \bibfield  {author} {\bibinfo {author} {\bibfnamefont {D.}~\bibnamefont
  {McDermott}}, \bibinfo {author} {\bibfnamefont {J.}~\bibnamefont {Amelang}},
  \bibinfo {author} {\bibfnamefont {L.~M.}\ \bibnamefont {Lopatina}}, \bibinfo
  {author} {\bibfnamefont {C.~J.~O.}\ \bibnamefont {Reichhardt}}, \ and\
  \bibinfo {author} {\bibfnamefont {C.}~\bibnamefont {Reichhardt}},\
  }\href@noop {} {\bibfield  {journal} {\bibinfo  {journal} {Soft Matter}\
  }\textbf {\bibinfo {volume} {9}},\ \bibinfo {pages} {4607} (\bibinfo {year}
  {2013})}\BibitemShut {NoStop}%
\bibitem [{\citenamefont {Zaidouny}\ \emph {et~al.}(2013)\citenamefont
  {Zaidouny}, \citenamefont {Bohlein}, \citenamefont {Roth},\ and\
  \citenamefont {Bechinger}}]{Zaidouny2013}%
  \BibitemOpen
  \bibfield  {author} {\bibinfo {author} {\bibfnamefont {L.}~\bibnamefont
  {Zaidouny}}, \bibinfo {author} {\bibfnamefont {T.}~\bibnamefont {Bohlein}},
  \bibinfo {author} {\bibfnamefont {R.}~\bibnamefont {Roth}}, \ and\ \bibinfo
  {author} {\bibfnamefont {C.}~\bibnamefont {Bechinger}},\ }\href@noop {}
  {\bibfield  {journal} {\bibinfo  {journal} {Soft Matter}\ }\textbf {\bibinfo
  {volume} {9}},\ \bibinfo {pages} {9230} (\bibinfo {year} {2013})}\BibitemShut
  {NoStop}%
\bibitem [{\citenamefont {Wang}\ \emph {et~al.}(2013)\citenamefont {Wang},
  \citenamefont {Tekic}, \citenamefont {Duan}, \citenamefont {Shao},\ and\
  \citenamefont {Yang}}]{Wang2013}%
  \BibitemOpen
  \bibfield  {author} {\bibinfo {author} {\bibfnamefont {C.~L.}\ \bibnamefont
  {Wang}}, \bibinfo {author} {\bibfnamefont {J.}~\bibnamefont {Tekic}},
  \bibinfo {author} {\bibfnamefont {W.~S.}\ \bibnamefont {Duan}}, \bibinfo
  {author} {\bibfnamefont {Z.~G.}\ \bibnamefont {Shao}}, \ and\ \bibinfo
  {author} {\bibfnamefont {L.}~\bibnamefont {Yang}},\ }\href@noop {} {\bibfield
   {journal} {\bibinfo  {journal} {Journal of Chemical Physics}\ }\textbf
  {\bibinfo {volume} {138}},\ \bibinfo {pages} {034307} (\bibinfo {year}
  {2013})}\BibitemShut {NoStop}%
\bibitem [{\citenamefont {Tekic}\ and\ \citenamefont
  {Mali}(2015)}]{TekicMali-book}%
  \BibitemOpen
  \bibfield  {author} {\bibinfo {author} {\bibfnamefont {J.}~\bibnamefont
  {Tekic}}\ and\ \bibinfo {author} {\bibfnamefont {P.}~\bibnamefont {Mali}},\
  }\href@noop {} {\emph {\bibinfo {title} {The AC Driven Frenkel-Kontorova
  Model}}}\ (\bibinfo  {publisher} {Faculty of Science},\ \bibinfo {address}
  {University of Novi Sad, Serbia},\ \bibinfo {year} {2015})\BibitemShut
  {NoStop}%
\bibitem [{\citenamefont {Pelton}\ \emph {et~al.}(2004)\citenamefont {Pelton},
  \citenamefont {Ladavac},\ and\ \citenamefont {Grier}}]{Pelton2004}%
  \BibitemOpen
  \bibfield  {author} {\bibinfo {author} {\bibfnamefont {M.}~\bibnamefont
  {Pelton}}, \bibinfo {author} {\bibfnamefont {K.}~\bibnamefont {Ladavac}}, \
  and\ \bibinfo {author} {\bibfnamefont {D.~G.}\ \bibnamefont {Grier}},\
  }\href@noop {} {\bibfield  {journal} {\bibinfo  {journal} {Physical Review
  E}\ }\textbf {\bibinfo {volume} {70}},\ \bibinfo {pages} {031108} (\bibinfo
  {year} {2004})}\BibitemShut {NoStop}%
\bibitem [{\citenamefont {Bohlein}\ \emph {et~al.}(2012)\citenamefont
  {Bohlein}, \citenamefont {Mikhael},\ and\ \citenamefont
  {Bechinger}}]{Bohlein2012}%
  \BibitemOpen
  \bibfield  {author} {\bibinfo {author} {\bibfnamefont {T.}~\bibnamefont
  {Bohlein}}, \bibinfo {author} {\bibfnamefont {J.}~\bibnamefont {Mikhael}}, \
  and\ \bibinfo {author} {\bibfnamefont {C.}~\bibnamefont {Bechinger}},\
  }\href@noop {} {\bibfield  {journal} {\bibinfo  {journal} {Nature Materials}\
  }\textbf {\bibinfo {volume} {11}},\ \bibinfo {pages} {126} (\bibinfo {year}
  {2012})}\BibitemShut {NoStop}%
\bibitem [{\citenamefont {Lindenberg}\ \emph {et~al.}(2005)\citenamefont
  {Lindenberg}, \citenamefont {Lacasta}, \citenamefont {Sancho},\ and\
  \citenamefont {Romero}}]{Lindenberg2005}%
  \BibitemOpen
  \bibfield  {author} {\bibinfo {author} {\bibfnamefont {K.}~\bibnamefont
  {Lindenberg}}, \bibinfo {author} {\bibfnamefont {A.~M.}\ \bibnamefont
  {Lacasta}}, \bibinfo {author} {\bibfnamefont {J.~M.}\ \bibnamefont {Sancho}},
  \ and\ \bibinfo {author} {\bibfnamefont {A.~H.}\ \bibnamefont {Romero}},\
  }\href@noop {} {\bibfield  {journal} {\bibinfo  {journal} {New Journal of
  Physics}\ }\textbf {\bibinfo {volume} {7}},\ \bibinfo {pages} {29} (\bibinfo
  {year} {2005})}\BibitemShut {NoStop}%
\bibitem [{\citenamefont {Shi}\ and\ \citenamefont {Ugaz}(2014)}]{Shi2014}%
  \BibitemOpen
  \bibfield  {author} {\bibinfo {author} {\bibfnamefont {N.}~\bibnamefont
  {Shi}}\ and\ \bibinfo {author} {\bibfnamefont {V.~M.}\ \bibnamefont {Ugaz}},\
  }\href@noop {} {\bibfield  {journal} {\bibinfo  {journal} {Physical Review
  E}\ }\textbf {\bibinfo {volume} {89}},\ \bibinfo {pages} {012138} (\bibinfo
  {year} {2014})}\BibitemShut {NoStop}%
\bibitem [{\citenamefont {Reguera}\ \emph {et~al.}(2002)\citenamefont
  {Reguera}, \citenamefont {Reimann}, \citenamefont {Hanggi},\ and\
  \citenamefont {Rubi}}]{Reguera2002}%
  \BibitemOpen
  \bibfield  {author} {\bibinfo {author} {\bibfnamefont {D.}~\bibnamefont
  {Reguera}}, \bibinfo {author} {\bibfnamefont {P.}~\bibnamefont {Reimann}},
  \bibinfo {author} {\bibfnamefont {P.}~\bibnamefont {Hanggi}}, \ and\ \bibinfo
  {author} {\bibfnamefont {J.~M.}\ \bibnamefont {Rubi}},\ }\href@noop {}
  {\bibfield  {journal} {\bibinfo  {journal} {Europhysics Letters}\ }\textbf
  {\bibinfo {volume} {57}},\ \bibinfo {pages} {644} (\bibinfo {year}
  {2002})}\BibitemShut {NoStop}%
\bibitem [{\citenamefont {Tekic}\ and\ \citenamefont {Hu}(2008)}]{Tekic2008}%
  \BibitemOpen
  \bibfield  {author} {\bibinfo {author} {\bibfnamefont {J.}~\bibnamefont
  {Tekic}}\ and\ \bibinfo {author} {\bibfnamefont {B.}~\bibnamefont {Hu}},\
  }\href@noop {} {\bibfield  {journal} {\bibinfo  {journal} {Physical Review
  B}\ }\textbf {\bibinfo {volume} {78}},\ \bibinfo {pages} {104305} (\bibinfo
  {year} {2008})}\BibitemShut {NoStop}%
\bibitem [{\citenamefont {Arzola}\ \emph {et~al.}(2013)\citenamefont {Arzola},
  \citenamefont {Volke-Sepúlveda},\ and\ \citenamefont {Mateos}}]{Arzola2013}%
  \BibitemOpen
  \bibfield  {author} {\bibinfo {author} {\bibfnamefont {A.~V.}\ \bibnamefont
  {Arzola}}, \bibinfo {author} {\bibfnamefont {K.}~\bibnamefont
  {Volke-Sepúlveda}}, \ and\ \bibinfo {author} {\bibfnamefont {J.~L.}\
  \bibnamefont {Mateos}},\ }\href@noop {} {\bibfield  {journal} {\bibinfo
  {journal} {Physical Review E}\ }\textbf {\bibinfo {volume} {87}},\ \bibinfo
  {pages} {062910} (\bibinfo {year} {2013})}\BibitemShut {NoStop}%
\bibitem [{\citenamefont {Herrera-Velarde}\ and\ \citenamefont
  {Castaneda-Priego}(2008)}]{Herrera2008}%
  \BibitemOpen
  \bibfield  {author} {\bibinfo {author} {\bibfnamefont {S.}~\bibnamefont
  {Herrera-Velarde}}\ and\ \bibinfo {author} {\bibfnamefont {R.}~\bibnamefont
  {Castaneda-Priego}},\ }\href@noop {} {\bibfield  {journal} {\bibinfo
  {journal} {Physical Review E}\ }\textbf {\bibinfo {volume} {77}},\ \bibinfo
  {pages} {041407} (\bibinfo {year} {2008})}\BibitemShut {NoStop}%
\bibitem [{\citenamefont {Lichtner}\ \emph {et~al.}(2012)\citenamefont
  {Lichtner}, \citenamefont {Pototsky},\ and\ \citenamefont
  {Klapp}}]{Lichtner2012}%
  \BibitemOpen
  \bibfield  {author} {\bibinfo {author} {\bibfnamefont {K.}~\bibnamefont
  {Lichtner}}, \bibinfo {author} {\bibfnamefont {A.}~\bibnamefont {Pototsky}},
  \ and\ \bibinfo {author} {\bibfnamefont {S.~H.~L.}\ \bibnamefont {Klapp}},\
  }\href@noop {} {\bibfield  {journal} {\bibinfo  {journal} {Physical Review
  E}\ }\textbf {\bibinfo {volume} {86}},\ \bibinfo {pages} {051405} (\bibinfo
  {year} {2012})}\BibitemShut {NoStop}%
\bibitem [{\citenamefont {Chen}\ and\ \citenamefont {Jiao}(2007)}]{Chen2007}%
  \BibitemOpen
  \bibfield  {author} {\bibinfo {author} {\bibfnamefont {J.~X.}\ \bibnamefont
  {Chen}}\ and\ \bibinfo {author} {\bibfnamefont {Z.~K.}\ \bibnamefont
  {Jiao}},\ }\href@noop {} {\bibfield  {journal} {\bibinfo  {journal} {Chinese
  Physics Letters}\ }\textbf {\bibinfo {volume} {24}},\ \bibinfo {pages} {1095}
  (\bibinfo {year} {2007})}\BibitemShut {NoStop}%
\bibitem [{\citenamefont {Straube}\ and\ \citenamefont
  {Tierno}(2013)}]{Straube2013}%
  \BibitemOpen
  \bibfield  {author} {\bibinfo {author} {\bibfnamefont {A.~V.}\ \bibnamefont
  {Straube}}\ and\ \bibinfo {author} {\bibfnamefont {P.}~\bibnamefont
  {Tierno}},\ }\href@noop {} {\bibfield  {journal} {\bibinfo  {journal}
  {Europhysics Letters}\ }\textbf {\bibinfo {volume} {103}},\ \bibinfo {pages}
  {28001} (\bibinfo {year} {2013})}\BibitemShut {NoStop}%
\bibitem [{\citenamefont {Song}\ \emph {et~al.}(2015)\citenamefont {Song},
  \citenamefont {Wang}, \citenamefont {Ren},\ and\ \citenamefont
  {Cao}}]{Song2015}%
  \BibitemOpen
  \bibfield  {author} {\bibinfo {author} {\bibfnamefont {K.~N.}\ \bibnamefont
  {Song}}, \bibinfo {author} {\bibfnamefont {H.~L.}\ \bibnamefont {Wang}},
  \bibinfo {author} {\bibfnamefont {J.}~\bibnamefont {Ren}}, \ and\ \bibinfo
  {author} {\bibfnamefont {Y.~G.}\ \bibnamefont {Cao}},\ }\href@noop {}
  {\bibfield  {journal} {\bibinfo  {journal} {Physica A}\ }\textbf {\bibinfo
  {volume} {417}},\ \bibinfo {pages} {102} (\bibinfo {year}
  {2015})}\BibitemShut {NoStop}%
\bibitem [{\citenamefont {Paronuzzi~Ticco}\ \emph {et~al.}(2016)\citenamefont
  {Paronuzzi~Ticco}, \citenamefont {Fornasier}, \citenamefont {Manini},
  \citenamefont {Santoro}, \citenamefont {Tosatti},\ and\ \citenamefont
  {Vanossi}}]{Paronuzzi2016}%
  \BibitemOpen
  \bibfield  {author} {\bibinfo {author} {\bibfnamefont {S.~V.}\ \bibnamefont
  {Paronuzzi~Ticco}}, \bibinfo {author} {\bibfnamefont {G.}~\bibnamefont
  {Fornasier}}, \bibinfo {author} {\bibfnamefont {N.}~\bibnamefont {Manini}},
  \bibinfo {author} {\bibfnamefont {G.~E.}\ \bibnamefont {Santoro}}, \bibinfo
  {author} {\bibfnamefont {E.}~\bibnamefont {Tosatti}}, \ and\ \bibinfo
  {author} {\bibfnamefont {A.}~\bibnamefont {Vanossi}},\ }\href@noop {}
  {\bibfield  {journal} {\bibinfo  {journal} {Journal of Physics: Condensed
  Matter}\ }\textbf {\bibinfo {volume} {28}},\ \bibinfo {pages} {134006}
  (\bibinfo {year} {2016})}\BibitemShut {NoStop}%
\bibitem [{\citenamefont {Juniper}\ \emph {et~al.}(2015)\citenamefont
  {Juniper}, \citenamefont {Straube}, \citenamefont {Besseling}, \citenamefont
  {Aarts},\ and\ \citenamefont {Dullens}}]{Juniper2015}%
  \BibitemOpen
  \bibfield  {author} {\bibinfo {author} {\bibfnamefont {M.~P.~N.}\
  \bibnamefont {Juniper}}, \bibinfo {author} {\bibfnamefont {A.~V.}\
  \bibnamefont {Straube}}, \bibinfo {author} {\bibfnamefont {R.}~\bibnamefont
  {Besseling}}, \bibinfo {author} {\bibfnamefont {D.~G. A.~L.}\ \bibnamefont
  {Aarts}}, \ and\ \bibinfo {author} {\bibfnamefont {R.~P.~A.}\ \bibnamefont
  {Dullens}},\ }\href@noop {} {\bibfield  {journal} {\bibinfo  {journal}
  {Nature Communications}\ }\textbf {\bibinfo {volume} {6}},\ \bibinfo {pages}
  {7187} (\bibinfo {year} {2015})}\BibitemShut {NoStop}%
\bibitem [{\citenamefont {Juniper}\ \emph {et~al.}(2012)\citenamefont
  {Juniper}, \citenamefont {Besseling}, \citenamefont {Aarts},\ and\
  \citenamefont {Dullens}}]{Juniper2012}%
  \BibitemOpen
  \bibfield  {author} {\bibinfo {author} {\bibfnamefont {M.~P.~N.}\
  \bibnamefont {Juniper}}, \bibinfo {author} {\bibfnamefont {R.}~\bibnamefont
  {Besseling}}, \bibinfo {author} {\bibfnamefont {D.~G. A.~L.}\ \bibnamefont
  {Aarts}}, \ and\ \bibinfo {author} {\bibfnamefont {R.~P.~A.}\ \bibnamefont
  {Dullens}},\ }\href@noop {} {\bibfield  {journal} {\bibinfo  {journal}
  {Optics Express}\ }\textbf {\bibinfo {volume} {27}},\ \bibinfo {pages}
  {28707} (\bibinfo {year} {2012})}\BibitemShut {NoStop}%
\bibitem [{\citenamefont {Juniper}\ \emph {et~al.}(2016)\citenamefont
  {Juniper}, \citenamefont {Straube}, \citenamefont {Aarts},\ and\
  \citenamefont {Dullens}}]{Juniper2016}%
  \BibitemOpen
  \bibfield  {author} {\bibinfo {author} {\bibfnamefont {M.~P.~N.}\
  \bibnamefont {Juniper}}, \bibinfo {author} {\bibfnamefont {A.~V.}\
  \bibnamefont {Straube}}, \bibinfo {author} {\bibfnamefont {D.~G. A.~L.}\
  \bibnamefont {Aarts}}, \ and\ \bibinfo {author} {\bibfnamefont {R.~P.~A.}\
  \bibnamefont {Dullens}},\ }\href@noop {} {\bibfield  {journal} {\bibinfo
  {journal} {Physical Review E}\ }\textbf {\bibinfo {volume} {93}},\ \bibinfo
  {pages} {012608} (\bibinfo {year} {2016})}\BibitemShut {NoStop}%
\bibitem [{\citenamefont {Adler}(1946)}]{Adler1946}%
  \BibitemOpen
  \bibfield  {author} {\bibinfo {author} {\bibfnamefont {R.}~\bibnamefont
  {Adler}},\ }\href@noop {} {\bibfield  {journal} {\bibinfo  {journal}
  {Proceedings of the IRE}\ }\textbf {\bibinfo {volume} {34}},\ \bibinfo
  {pages} {351} (\bibinfo {year} {1946})}\BibitemShut {NoStop}%
\bibitem [{\citenamefont {Goldstein}\ \emph {et~al.}(2009)\citenamefont
  {Goldstein}, \citenamefont {Polin},\ and\ \citenamefont
  {Tuval}}]{Goldstein2009}%
  \BibitemOpen
  \bibfield  {author} {\bibinfo {author} {\bibfnamefont {R.~E.}\ \bibnamefont
  {Goldstein}}, \bibinfo {author} {\bibfnamefont {M.}~\bibnamefont {Polin}}, \
  and\ \bibinfo {author} {\bibfnamefont {I.}~\bibnamefont {Tuval}},\
  }\href@noop {} {\bibfield  {journal} {\bibinfo  {journal} {Physical Review
  Letters}\ }\textbf {\bibinfo {volume} {103}},\ \bibinfo {pages} {168103}
  (\bibinfo {year} {2009})}\BibitemShut {NoStop}%
\bibitem [{\citenamefont {Reichhardt}\ \emph {et~al.}(2000)\citenamefont
  {Reichhardt}, \citenamefont {Scalettar}, \citenamefont {Zimanyi},\ and\
  \citenamefont {Gronbech-Jensen}}]{Reichhardt2000}%
  \BibitemOpen
  \bibfield  {author} {\bibinfo {author} {\bibfnamefont {C.}~\bibnamefont
  {Reichhardt}}, \bibinfo {author} {\bibfnamefont {R.~T.}\ \bibnamefont
  {Scalettar}}, \bibinfo {author} {\bibfnamefont {G.~T.}\ \bibnamefont
  {Zimanyi}}, \ and\ \bibinfo {author} {\bibfnamefont {N.}~\bibnamefont
  {Gronbech-Jensen}},\ }\href@noop {} {\bibfield  {journal} {\bibinfo
  {journal} {Physica C}\ }\textbf {\bibinfo {volume} {332}},\ \bibinfo {pages}
  {1} (\bibinfo {year} {2000})}\BibitemShut {NoStop}%
\bibitem [{\citenamefont {Marconi}\ and\ \citenamefont
  {Tarazona}(1999)}]{Tarazona_Marconi_JCP}%
  \BibitemOpen
  \bibfield  {author} {\bibinfo {author} {\bibfnamefont {U.~M.~B.}\
  \bibnamefont {Marconi}}\ and\ \bibinfo {author} {\bibfnamefont
  {P.}~\bibnamefont {Tarazona}},\ }\href@noop {} {\bibfield  {journal}
  {\bibinfo  {journal} {The Journal of Chemical Physics}\ }\textbf {\bibinfo
  {volume} {110}},\ \bibinfo {pages} {8032} (\bibinfo {year}
  {1999})}\BibitemShut {NoStop}%
\bibitem [{\citenamefont {Archer}\ and\ \citenamefont
  {Evans}(2004)}]{Archer_Evans_JCP}%
  \BibitemOpen
  \bibfield  {author} {\bibinfo {author} {\bibfnamefont {A.~J.}\ \bibnamefont
  {Archer}}\ and\ \bibinfo {author} {\bibfnamefont {R.}~\bibnamefont {Evans}},\
  }\href@noop {} {\bibfield  {journal} {\bibinfo  {journal} {The Journal of
  Chemical Physics}\ }\textbf {\bibinfo {volume} {121}},\ \bibinfo {pages}
  {4246} (\bibinfo {year} {2004})}\BibitemShut {NoStop}%
\bibitem [{\citenamefont {Espa\~{n}ol}\ and\ \citenamefont
  {L\"{o}wen}(2009)}]{Espanol_Lowen_JCP_2009}%
  \BibitemOpen
  \bibfield  {author} {\bibinfo {author} {\bibfnamefont {P.}~\bibnamefont
  {Espa\~{n}ol}}\ and\ \bibinfo {author} {\bibfnamefont {H.}~\bibnamefont
  {L\"{o}wen}},\ }\href@noop {} {\bibfield  {journal} {\bibinfo  {journal} {The
  Journal of Chemical Physics}\ }\textbf {\bibinfo {volume} {131}},\ \bibinfo
  {pages} {244101} (\bibinfo {year} {2009})}\BibitemShut {NoStop}%
\bibitem [{\citenamefont {Guyer}\ \emph {et~al.}(2009)\citenamefont {Guyer},
  \citenamefont {Wheeler},\ and\ \citenamefont {Warren}}]{FiPy:2009}%
  \BibitemOpen
  \bibfield  {author} {\bibinfo {author} {\bibfnamefont {J.~E.}\ \bibnamefont
  {Guyer}}, \bibinfo {author} {\bibfnamefont {D.}~\bibnamefont {Wheeler}}, \
  and\ \bibinfo {author} {\bibfnamefont {J.~A.}\ \bibnamefont {Warren}},\
  }\href@noop {} {\bibfield  {journal} {\bibinfo  {journal} {Computing in
  Science \& Engineering}\ }\textbf {\bibinfo {volume} {11}},\ \bibinfo {pages}
  {6} (\bibinfo {year} {2009})}\BibitemShut {NoStop}%
\bibitem [{\citenamefont {Guo}\ \emph {et~al.}(2014)\citenamefont {Guo},
  \citenamefont {Du},\ and\ \citenamefont {Mei}}]{Guo2014}%
  \BibitemOpen
  \bibfield  {author} {\bibinfo {author} {\bibfnamefont {W.}~\bibnamefont
  {Guo}}, \bibinfo {author} {\bibfnamefont {L.-C.}\ \bibnamefont {Du}}, \ and\
  \bibinfo {author} {\bibfnamefont {D.-C.}\ \bibnamefont {Mei}},\ }\href@noop
  {} {\bibfield  {journal} {\bibinfo  {journal} {J. Stat. Mech.}\ ,\ \bibinfo
  {pages} {P04025}} (\bibinfo {year} {2014})}\BibitemShut {NoStop}%
\bibitem [{\citenamefont {Marchenko}\ and\ \citenamefont
  {Marchenko}(2012)}]{Marchenko2012}%
  \BibitemOpen
  \bibfield  {author} {\bibinfo {author} {\bibfnamefont {I.~G.}\ \bibnamefont
  {Marchenko}}\ and\ \bibinfo {author} {\bibfnamefont {I.~I.}\ \bibnamefont
  {Marchenko}},\ }\href@noop {} {\bibfield  {journal} {\bibinfo  {journal}
  {JETP Lett.}\ }\textbf {\bibinfo {volume} {95}},\ \bibinfo {pages} {137}
  (\bibinfo {year} {2012})}\BibitemShut {NoStop}%
\bibitem [{\citenamefont {Saikia}\ and\ \citenamefont
  {Mahato}(2009)}]{Saikia2009}%
  \BibitemOpen
  \bibfield  {author} {\bibinfo {author} {\bibfnamefont {S.}~\bibnamefont
  {Saikia}}\ and\ \bibinfo {author} {\bibfnamefont {M.~C.}\ \bibnamefont
  {Mahato}},\ }\href@noop {} {\bibfield  {journal} {\bibinfo  {journal} {Phys.
  Rev. E}\ }\textbf {\bibinfo {volume} {80}},\ \bibinfo {pages} {062102}
  (\bibinfo {year} {2009})}\BibitemShut {NoStop}%
\bibitem [{\citenamefont {Lindenberg}\ \emph {et~al.}(2007)\citenamefont
  {Lindenberg}, \citenamefont {Sancho}, \citenamefont {Lacasta},\ and\
  \citenamefont {Sokolov}}]{Lindenberg2007}%
  \BibitemOpen
  \bibfield  {author} {\bibinfo {author} {\bibfnamefont {K.}~\bibnamefont
  {Lindenberg}}, \bibinfo {author} {\bibfnamefont {J.~M.}\ \bibnamefont
  {Sancho}}, \bibinfo {author} {\bibfnamefont {A.~M.}\ \bibnamefont {Lacasta}},
  \ and\ \bibinfo {author} {\bibfnamefont {I.~M.}\ \bibnamefont {Sokolov}},\
  }\href@noop {} {\bibfield  {journal} {\bibinfo  {journal} {Phys. Rev. Lett.}\
  }\textbf {\bibinfo {volume} {98}},\ \bibinfo {pages} {020602} (\bibinfo
  {year} {2007})}\BibitemShut {NoStop}%
\bibitem [{\citenamefont {Emary}\ and\ \citenamefont
  {Klapp}(2012)}]{Emary2012}%
  \BibitemOpen
  \bibfield  {author} {\bibinfo {author} {\bibfnamefont {R.}~\bibnamefont
  {Emary}, \bibfnamefont {C.and~Gernert}}\ and\ \bibinfo {author}
  {\bibfnamefont {S.~H.~L.}\ \bibnamefont {Klapp}},\ }\href@noop {} {\bibfield
  {journal} {\bibinfo  {journal} {Phys. Rev. E}\ }\textbf {\bibinfo {volume}
  {86}},\ \bibinfo {pages} {061135} (\bibinfo {year} {2012})}\BibitemShut
  {NoStop}%
\bibitem [{\citenamefont {Wiesenfeld}\ and\ \citenamefont
  {Satija}(1987)}]{Wiesenfeld1987}%
  \BibitemOpen
  \bibfield  {author} {\bibinfo {author} {\bibfnamefont {K.}~\bibnamefont
  {Wiesenfeld}}\ and\ \bibinfo {author} {\bibfnamefont {I.}~\bibnamefont
  {Satija}},\ }\href@noop {} {\bibfield  {journal} {\bibinfo  {journal} {Phys.
  Rev. B}\ }\textbf {\bibinfo {volume} {36}},\ \bibinfo {pages} {2483}
  (\bibinfo {year} {1987})}\BibitemShut {NoStop}%
\bibitem [{\citenamefont {Crommie}\ \emph {et~al.}(1991)\citenamefont
  {Crommie}, \citenamefont {Craig}, \citenamefont {Sherwin},\ and\
  \citenamefont {Zettl}}]{Crommie1991}%
  \BibitemOpen
  \bibfield  {author} {\bibinfo {author} {\bibfnamefont {M.~F.}\ \bibnamefont
  {Crommie}}, \bibinfo {author} {\bibfnamefont {K.}~\bibnamefont {Craig}},
  \bibinfo {author} {\bibfnamefont {M.~S.}\ \bibnamefont {Sherwin}}, \ and\
  \bibinfo {author} {\bibfnamefont {A.}~\bibnamefont {Zettl}},\ }\href@noop {}
  {\bibfield  {journal} {\bibinfo  {journal} {Phys. Rev. B}\ }\textbf {\bibinfo
  {volume} {43}},\ \bibinfo {pages} {13699} (\bibinfo {year}
  {1991})}\BibitemShut {NoStop}%
\bibitem [{\citenamefont {Hu}\ and\ \citenamefont {Tekic}(2007)}]{Hu2007}%
  \BibitemOpen
  \bibfield  {author} {\bibinfo {author} {\bibfnamefont {B.}~\bibnamefont
  {Hu}}\ and\ \bibinfo {author} {\bibfnamefont {J.}~\bibnamefont {Tekic}},\
  }\href@noop {} {\bibfield  {journal} {\bibinfo  {journal} {Physical Review
  E}\ }\textbf {\bibinfo {volume} {75}},\ \bibinfo {pages} {056608} (\bibinfo
  {year} {2007})}\BibitemShut {NoStop}%
\bibitem [{\citenamefont {Cotteverte}\ \emph {et~al.}(1994)\citenamefont
  {Cotteverte}, \citenamefont {Bretenaker},\ and\ \citenamefont
  {Lefloch}}]{Cotteverte1994}%
  \BibitemOpen
  \bibfield  {author} {\bibinfo {author} {\bibfnamefont {J.~C.}\ \bibnamefont
  {Cotteverte}}, \bibinfo {author} {\bibfnamefont {F.}~\bibnamefont
  {Bretenaker}}, \ and\ \bibinfo {author} {\bibfnamefont {A.}~\bibnamefont
  {Lefloch}},\ }\href@noop {} {\bibfield  {journal} {\bibinfo  {journal}
  {Physical Review A}\ }\textbf {\bibinfo {volume} {49}},\ \bibinfo {pages}
  {2868} (\bibinfo {year} {1994})}\BibitemShut {NoStop}%
\bibitem [{\citenamefont {Chow}\ \emph {et~al.}(1985)\citenamefont {Chow},
  \citenamefont {Geabanacloche}, \citenamefont {Pedrotti}, \citenamefont
  {Sanders}, \citenamefont {Schleich},\ and\ \citenamefont
  {Scully}}]{Chow1985}%
  \BibitemOpen
  \bibfield  {author} {\bibinfo {author} {\bibfnamefont {W.~W.}\ \bibnamefont
  {Chow}}, \bibinfo {author} {\bibfnamefont {J.}~\bibnamefont {Geabanacloche}},
  \bibinfo {author} {\bibfnamefont {L.~M.}\ \bibnamefont {Pedrotti}}, \bibinfo
  {author} {\bibfnamefont {V.~E.}\ \bibnamefont {Sanders}}, \bibinfo {author}
  {\bibfnamefont {W.}~\bibnamefont {Schleich}}, \ and\ \bibinfo {author}
  {\bibfnamefont {M.~O.}\ \bibnamefont {Scully}},\ }\href@noop {} {\bibfield
  {journal} {\bibinfo  {journal} {Reviews of Modern Physics}\ }\textbf
  {\bibinfo {volume} {57}},\ \bibinfo {pages} {61} (\bibinfo {year}
  {1985})}\BibitemShut {NoStop}%
\end{thebibliography}%

\end{document}